\documentclass{jpp}
\usepackage{graphicx, xcolor}
\usepackage{epstopdf, epsfig}
\usepackage{hyperref}
\usepackage{amsmath,amsfonts,amssymb} 
\usepackage{subfigure} 
\usepackage{natbib}
\usepackage{float}
\usepackage{dblfloatfix}
\usepackage{tabularx}

\newcommand{\be}{\begin{equation}} 
\newcommand{\ee}{\end{equation}} 
\newcommand{\bea}{\begin{eqnarray}} 
\newcommand{\eea}{\end{eqnarray}} 
\newcommand{\vomega}{\mbox{\boldmath $\omega$}}

\shorttitle{Kurtosis--skewness  relations  in  fast dynamo regimes in MHD}
\shortauthor{Yannick Ponty, H\'el\`ene Politano and Annick Pouquet}
\title{Intermittency assessed through a model \\
of kurtosis-skewness relation in \\
  MHD in  fast dynamo regimes}
\author{Yannick Ponty\aff{1}\corresp{\email{yannick.ponty@oca.eu}}, H\'el\`ene Politano\aff{2} and Annick Pouquet\aff{3}}
\affiliation{\aff{1}Université Côte d’Azur, CNRS, Observatoire de la Côte d’Azur, Laboratoire Lagrange, France 
\aff{2}Université Côte d’Azur, CNRS, Laboratoire J. A. Dieudonné, France 
\aff{3}National Center for Atmospheric Research, P.O.~Box 3000, Boulder, CO 80307, USA 
}
\begin{document}
\maketitle
\begin{abstract}
Intermittency as it occurs in  fast dynamos in the MHD framework is evaluated through the examination of relations between normalized moments at third order (skewness $S$) and fourth order (kurtosis $K$) for both the velocity and magnetic field, and for their  local dissipations. As investigated by several authors in various physical contexts such as fusion plasmas (\cite{krommes_08}), climate evolution 
 (\cite{sura_08}), fluid turbulence or rotating stratified flows (\cite{pouquet_23a}), approximate parabolic $K(S)\sim S^\alpha$ laws  emerge whose origin may be related to the applicability of intermittency models  to their dynamics. The results analyzed herein are obtained through direct numerical simulations of MHD flows for both Taylor-Green and  Beltrami ABC forcing at moderate Reynolds numbers, and for up to $3.14 \times 10^5$ turn-over times. We observe for the dissipation $0.2 \lesssim \alpha \lesssim 3.0$, an evaluation that varies with the field, the forcing, and when filtering for high-skewness intermittent structures. When using the \citet{she_94} intermittency model, one can compute   $\alpha$ analytically; we then find $\alpha \approx 2.5$, clearly differing  from a (strict) parabolic scaling, a result consistent with the numerical data. 
\end{abstract}

 \section{Introduction}  \label{S:I}   

One striking property of turbulent flows is their lack of predictability, as well as their intermittency, associated with the presence of intense and isolated patterns at small scales, such as vortex filaments and current sheets, or coherent structures at large scale. Such extreme events can be assessed through their probability distribution functions (PDFs), and thus through their normalized moments such as the  skewness and kurtosis (definitions are given in \S \ref{S:E}).  These intense structures have been identified in many experiments, observations and direct numerical simulations (see for recent reviews {\it e.g.} \citet{matthaeus_15, yeung_15, chen_16, camporeale_18p,  ergun_20, schekochihin_22}). The ensuing dissipation is found in a reduced volume  of the system, in both neutral fluids (\cite{bradshaw_19, rafner_21, buaria_22p}) and MHD (see {\it e.g.} 
\cite{politano_95pp, meneguzzi_96, mallet_17j, zhdankin_17, adhikari_20}). This physical intermittency volume  can be in fact smaller than for fully developed turbulence (FDT),
{as shown explicitly} in the presence of gravity  waves (\cite{marino_22}).
{It is also found  in the ocean \citep{vanharen_16g}, as well as in clear air turbulence, the origin of which remains rather ill-understood although Kelvin-Helmholtz and shear instabilities are likely to be the culprit \citep{rodriguez_23}, as for the other examples given here.} 
 Similar observations are documented for plasma disruptions.

Intermittency can be characterized in many ways, such as when evaluating anomalous exponents of structure functions. Perhaps more directly, one can assess when and where normalized third and fourth-order moments, skewness $S$ and kurtosis $K$, differ from their Gaussian value.  There  is a long history of such measurements; for example, both skewness and kurtosis have been used to map a flow, such as in meandering jets in the ocean (\cite{hughes_10}), or in climate data reanalysis (\cite{petoukhov_08}).

One way to condense the data further is to look for $K(S)$ relations, often found to be close to parabolic in a variety of contexts (see  \citet{pouquet_23a} and references therein for a recent review), {\it e.g.} for the Navier-Stokes (NS) equations, in presence or not of stratification and/or rotation, as relevant to the atmosphere, the ocean and climate (\cite{lenschow_94, sura_08}). 
Such quasi-parabolic laws were also found in laboratory and astrophysical plasmas  (see for example \cite{labit_07, krommes_08, sattin_09, garcia_12, guszejnov_13, mezaoui_14, furno_15,  miranda_18}). 
More recently, \cite{sladkomedova_23}  analysed the intermittency of the density in MAST (Mega Ampere Spherical Tokamak) plasmas, and found that the data agrees with the $K(S)$ model  given by \cite{garcia_12} (see also \cite{guszejnov_13, losada_23})\footnote{The model is based on moments up to fourth order  as well as on an estimate of the number of the intermittent events that are observed simultaneously.}.

Detailed knowledge of  intermittency in plasmas and turbulent systems in general may lead to a better understanding of their PDFs and of the dissipation mechanisms at play. It is in this context that we want to examine in this paper the intermittent properties of MHD through the  possible relationship between excess kurtosis and skewness for the velocity and magnetic field, as well as for their local dissipation. MHD has, of course, many different regimes, and we concentrate here on one subset, namely that of the fast dynamo in its nonlinear phase and at moderate Reynolds number for which  long runs are available, up to in excess of $10^5 \tau_{nl}$, 
{where 
\be \tau_{nl}=L_{int}/V_{rms} \label{TNL} \ee
 is the turn-over time of the turbulence based on the large-scale velocity $V_{rms}$ and on the integral length scale $L_{int}$. In the next section  
are written the equations and definitions needed for our analysis of  numerical MHD data for a fast dynamo regime, as well as information on the direct numerical simulations (DNSs) employed herein. In  \S \ref{S:R},  we analyze the numerical results, and  in \S \ref{S:KS} we give an interpretation of them within a specific model of intermittency. The last section presents a discussion and our  conclusions. 
}

%
\section{Equations and numerical set-up} \label{S:E} 
\subsection{Equations and definitions}

 The equations for MHD in the incompressible case are written as usual as:
 \bea
[ \partial_t + {\bf u} \cdot \nabla ]{\bf u}  \equiv 
D_t {\bf u} &=& -\nabla P^\star +{\bf b} \cdot \nabla {\bf b} + \nu \Delta {\bf u} +{\bf F}_V   , \\  \label{mhd1}  
 [\partial_t + {\bf u} \cdot  \nabla ] {\bf b} \equiv D_t {\bf b} &=&  {\bf b} \cdot \nabla {\bf u} + \eta \Delta {\bf b}    , \label{mhd2} 
  \eea
together with $\nabla \cdot {\bf u}=0, \nabla \cdot {\bf b}=0$;
 ${\bf u}, {\bf b}$ are the velocity and magnetic field (in Alfv\'en velocity units), $P^\star$ is the total pressure, $\nu, \eta$ are the viscosity and magnetic diffusivity, and ${\bf F}_V$ is a forcing term. We will use in this paper two types of forcing.
The first one is the ABC (Arnold-Beltrami-Childress) forcing, which is a superposition of Beltrami vortices and thus fully helical; 
 it is defined as:
\be
{\bf f}_{\rm ABC} = \left[ \cos y +  \sin z \right] {\bf \hat{x}} + \left[\sin x +  \cos z \right] {\bf \hat{y}} + \left[ \cos x + \sin y \right] {\bf \hat{z}} .
\label{eq:ABC} \ee
The ABC flow is an eigenfunction of the curl, and  it is an exact solution of the Euler equations; thus,  for large enough viscosity, it is stable but turbulence develops as the Reynolds number $R_V$ increases. 
We also take the Taylor-Green (TG) forcing written as:
\be
{\bf f}_{\rm TG} = f_0^{tg} \left\{ \left[ \sin x \ \cos y \ \cos z \right] {\bf \hat x} -
                                                  \left[\cos x ~ \sin y \ \cos z \right] {\bf \hat y } + 0 ~ {\bf \hat z} 
                                          \right\}  \ ,
 \label{eq:TG}   \ee
with $f_0^{tg}=3$. This forcing is globally non-helical,  
but it can be viewed in fact as an assembly of helical patches of both signs and varied intensities. 

We solved numerically our MHD system in a fully periodic box using a classic pseudo-spectral solver, involving a $2/3$ dealiasing technique with a parallel CPU-MPI code (CUBBY; \cite{ponty_05}). With these two forcings, we compute four simulations altogether, for up to hundreds of thousands of eddy turn over times, and we record the same number of snapshots for the six three-dimensional  field components. Some of the data and a few of the statistical properties of these runs are given in Table 1. 

 \begin{table}   
  \begin{center}
    \begin{tabular}{ccccccccccc}    
      \hline
      Run -- & $n_p$  & $\nu$ & $R_V$ &  $R_\lambda$&  $R_N$ &  $T_{max}$ & $\tau_{nl}$ & $R_M$  & $P_m$  & $r$  \\
      \hline 
      TG1 -- & $64$ & 0.07 & 65.5  & 25.5 & 2.98  & $50\times 10^3$  & 2.23  &  111  &  0.7 & 1.05   \\ 
      \hline
      TG3 -- & $64$ & 0.1  & 60.  & 24.3 & 3.35  & $130\times 10^3$  & 1.86  &  430  & 3.3  &  5.26   \\ 
      \hline
      ABC1 -- & $64$ & 0.16  & 210. & 45.8 &  2.4 &  $60\times 10^3$   & 1.14  &  32  &  1.45 &  1.18    \\
      \hline
      ABC2 -- & $64$ & 0.2   & 175. & 41.   &  1.4  & $314.254\times 10^3$ &  1.12  &  21  &  1.42 & 1.21   \\
      \hline
      \end{tabular}  
    \end{center}
  \label{t:1} 
  \caption{
  Characteristics of the runs, with the linear resolution $n_p$ of the cubic grid, $\nu$ the viscosity, the Reynolds number $R_V= L_{int} V_{rms}/\nu$, and Taylor Reynolds number $R_\lambda=\lambda V_{rms}/\nu$;  $R_N=  \eta_v n_p/3$ is the so-called Kaneda {criterion} based on the resolution in terms of the Kolmogorov dissipation length $\eta_v$.  
  TG denotes Taylor-Green runs, and ABC denotes ABC runs (see text). 
  For each run, $T_{max}$ is its total duration, with $\tau_{nl}$ in these units between 1 and 2.
   For the magnetic variables, we give the magnetic Reynolds $R_M= L_{int} V_{rms}/\eta$, the magnetic Prandtl number $P_m=\nu/\eta$ and  $r=R_M/R_M^C$ the ratio of the magnetic Reynolds number to the (approximate) critical value for the threshold of the dynamo. All these non-dimensional numbers need different definition of scales, like the integral scale  $L_{int}=2 \pi \sum E(k)/k/\sum E(k)$ defined using the isotropic energy spectrum computed along the simulation at each wave number $k$, the Taylor scale  $\lambda=\sqrt{10}\eta_v R_V^{1/4}=\sqrt{10}L_{int}R_V^{-1/2}$ in the inertial range, and the Kolmogorov scale 
  $\eta_v= \left[\frac{\epsilon_v}{\nu^3}\right]^{-1/4} = L_{int}R_V^{-3/4}$ at the onset of the dissipation range. We need also one characteristic velocity which is usually taken as the root mean square of the kinetic energy $V_{rms}= \sqrt{2 \sum E(k)}$. The non-linear time is taken as $\tau_{nl}= L_{int}/V_{rms}$. All the scales and velocity are averaged in time as the simulations develop.     
  }
\end{table}

\subsection{{Brief}  description of the runs}

The dynamo problem, concerning the generation of magnetic fields at both large and small scales, is a long-standing topic (see \cite{brandenburg_05, rincon_dynamo_2019} for thorough recent reviews).
 In the context of this paper, we analyze four simulation runs, focusing on the turbulent dynamo regime.
In these simulations, the (fast) dynamo is triggered when the magnetic Reynolds number $R_M$ exceeds a threshold that depends on the magnetic Prandtl number $P_m=\nu/\eta$  (\cite{ponty_05}), as seen in run TG3. Sub-critical dynamos can also be observed, where magnetically-induced changes to the velocity field play a role, such as in run TG1 which is  close to the onset of dynamo action, and living on the sub-critical branch (\cite{ponty_07p}).
It is also worth noting that the Lorenz force's feedback on the velocity field can influence the so-called "on-off" intermittency of the dynamo, as explored in detail by \cite{alexakis_08} in the context of the ABC  runs ABC1 and ABC2. These runs are notable for their short {\it off} phases, during which the magnetic field becomes weak enough to revert the system to a hydrodynamic state. The dynamo comes back quickly, with a return to an  MHD equilibrium (see Fig. \ref{f:1} below).
We thus finally have  two different types of dynamo turbulence, with a large amount of fluctuations which are analyzed in the next sections,  {examining} the behavior of the  third and fourth-order normalized moment statistics using the full-space temporal field data.

\section{Analysis of the results} \label{S:R}
\subsection{Field gradient tensors, skewness and kurtosis}

Concerning the data points for measurements, which must be statistically independent, they are taken approximately every $\tau_{nl}$; 
we recall that measurement errors go as $\sqrt{6/N_s}$ with $N_s$ the number of  independent data points (see {\it e.g.} \cite{sura_08}). 
Large samples are needed also because a parabolic fit is quite sensitive to  extreme values. Note  that it is shown in \cite{wan_10}, in the context of two-dimensional (2D) MHD turbulence, that an estimate of the kurtosis at small scales requires, in the framework of the DNSs analyzed in that paper, that the (Kolmogorov) dissipation scale be at least twice as large as the cut-off $k_{max}=n_p/3$; this condition is well fulfilled by the runs of Table 1 (see parameter $R_N$).

We analyze the data using the point-wise rates of dissipation of the kinetic and magnetic energy, 
$ \epsilon_v({\bf x}),  \epsilon_m({\bf x})$.
They are expressed  in terms of the symmetric part of the velocity gradient tensor, namely the strain tensor $S^v_{ij}$:
\be 
S^v_{ij}({\bf x}) = \frac{\partial_j u_i ({\bf x})+ \partial_i u_j({\bf x})}{2} \ , \ 
\epsilon_v({\bf x}) = {2 \nu} \Sigma_{ij} S^v_{ij}({\bf x}) S^v_{ij}({\bf x}) \ ,
\ee 
and of the magnetic current density, {\it viz.}  $\epsilon_m({\bf x}) =  {\eta} {\bf j}^2({\bf x})$. 
For completeness,  we also define the symmetric part of the magnetic field gradient tensor, namely:
\be
S^m_{ij}({\bf x}) =  \frac{\partial_j b_i ({\bf x})+ \partial_i b_j({\bf x})}{2} \ , 
\ \sigma_m({\bf x}) = \Sigma_{ij} S^m_{ij}({\bf x}) S^m_{ij}({\bf x}) \ .
\label{BB} \ee
 Note that $S^m_{ij}$ is a pseudo (axial) tensor.  The standard expressions for the  integrated (space-averaged) kinetic, {magnetic and total energy dissipation rates,
 can respectively be written as:}
\be
\epsilon_V= \nu\left< { |\vomega}|^2 \right> \ , \  \epsilon_M = \ \eta \left< |{\bf j}^2| \right>   \ ,  \  
 \epsilon_T=\epsilon_V + \epsilon_M \ ,
\ee
with $\vomega=\nabla \times {\bf u}$ the vorticity. 
Finally, the skewness and excess kurtosis are written below for a scalar field $f$, with $S_f=0, K_f=0$ for a Gaussian distribution, with the kurtosis (or flatness) being defined as $F_f =K_f+3$:
\be S_f = \left<f^3 \right>/\left<f^2 \right>^{3/2},\  K_f = \left<f^4 \right>/\left<f^2 \right>^2-3\ .
\label{SK} \ee

 \subsection{Some models for $K(S)$ relations} \label{SS:MOD}
A  $K(S)$ parabolic law has been derived explicitly in \cite{sura_08} for a model of oceanic sea-surface temperature anomalies (SST), based on the dynamics of a specific linear Langevin model with both additive and multiplicative noises.
  Analyzing the corresponding Fokker-Planck equation for the stationary PDF, these authors can show analytically that in the limit of weak multiplicative noise,  {one has} $K(S)= 3S^2/2$.
Multiple other studies show the plausibility of a Langevin model for  parabolic $K(S)$ behaviors in different contexts as exemplified {\it e.g.} in \cite{hasselmann_62-1, sattin_09} 
{(see also \citet{pouquet_24} for  nonlinear Langevin models).}
 {Note that, in}  a Langevin equation, in a sense, one  is getting rid of the closure problem for turbulent motions since it is linear, with the complex nonlinear small-scale dynamics bundled up in a rapid stochastic forcing with an assumption of (mostly) local interactions among these fast motions.

{In fact,} models in the {framework}  of fusion plasmas have also been written, for example in the context of magnetically-confined experiments. 
In \citet{garcia_12}, the dynamics, as for dissipation events, is viewed as a random sequence of bursts as opposed to a quasi-continuum. These bursts are occurring independently and following a Poisson process. When taking for the shape of these bursts a sharp rise and a slow exponential decay, one can  compute $K(S)$ relations which, for an exponential distribution of burst amplitudes, becomes $K(S)=3/2(S^2-1)$.

\begin{figure*}  
 \includegraphics[width=7cm, height=5cm]{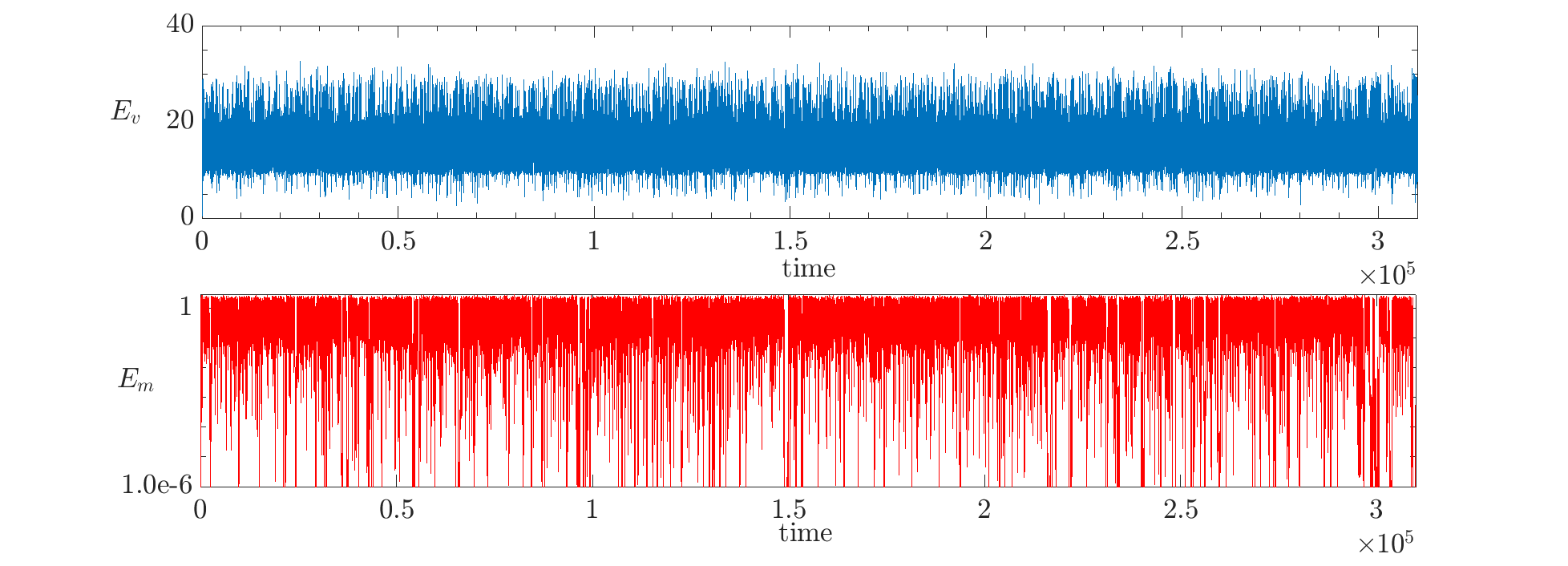}  
 \includegraphics[width=7cm, height=5cm]{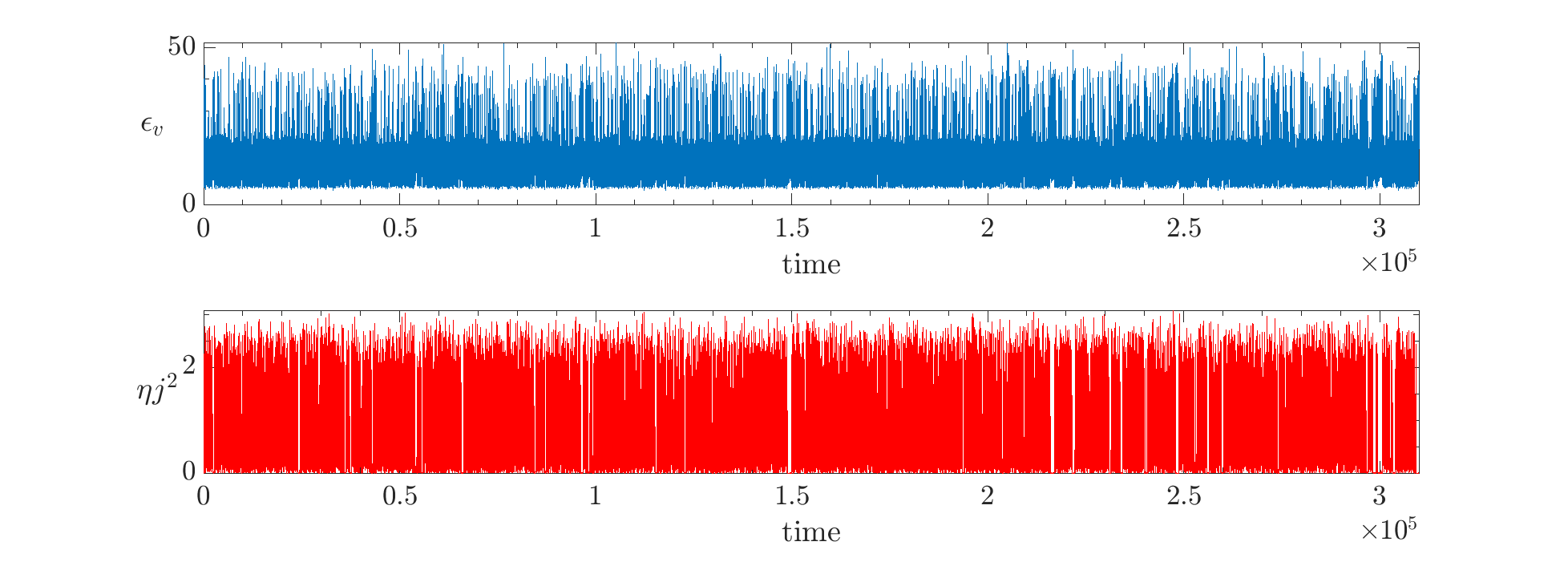}  
 \includegraphics[width=7cm, height=4cm]{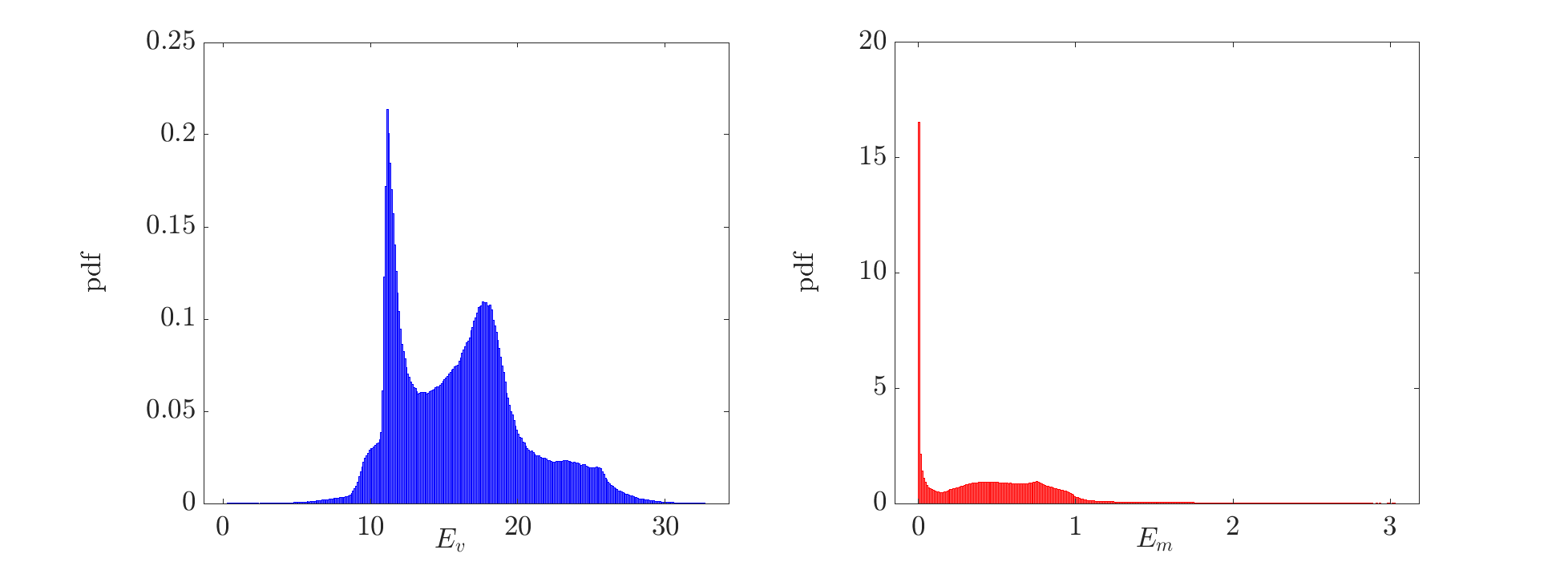}
 \includegraphics[width=7cm, height=4cm]{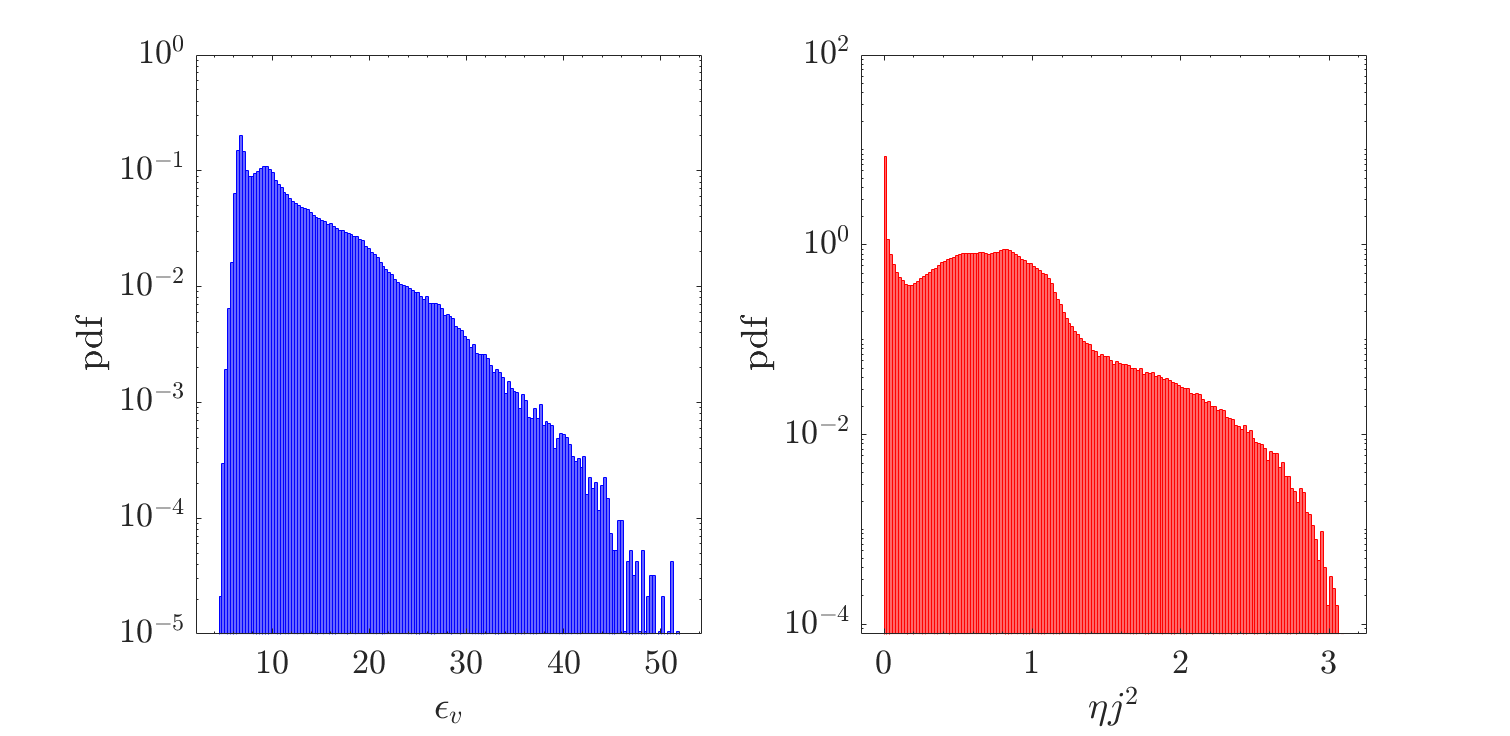}
 \caption{Run ABC2:  Temporal evolutions,  { with kinetic variables in blue and magnetic ones in red.
 Top, left and right: kinetic  energy and its dissipation, $\epsilon_v$.
 Middle, left and right: magnetic energy and its dissipation, $\epsilon_m=\eta {\bf j}^2$. 
 }
 Note the different units on  axes, and the lin-log scale for  magnetic energy. 
 Bottom: Probability density functions of the kinetic  
 {and magnetic energies (two left plots), and  of their dissipation (two right plots), with lin-log plots used for the latter.}
}  \label{f:1} 
\end{figure*}  

\begin{figure*}  
  \includegraphics[width=7cm, height=5cm]{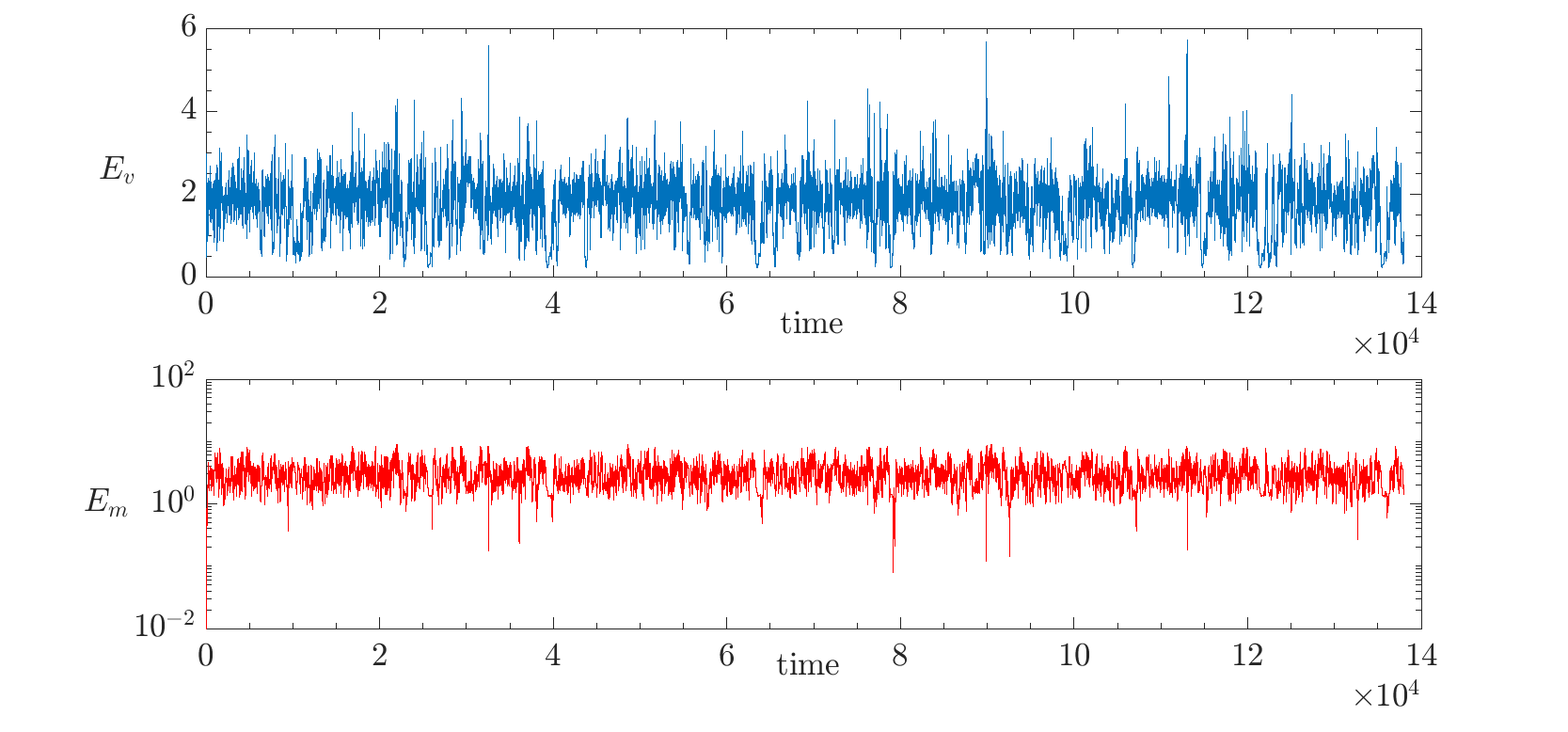} 
   \includegraphics[width=7cm, height=5cm]{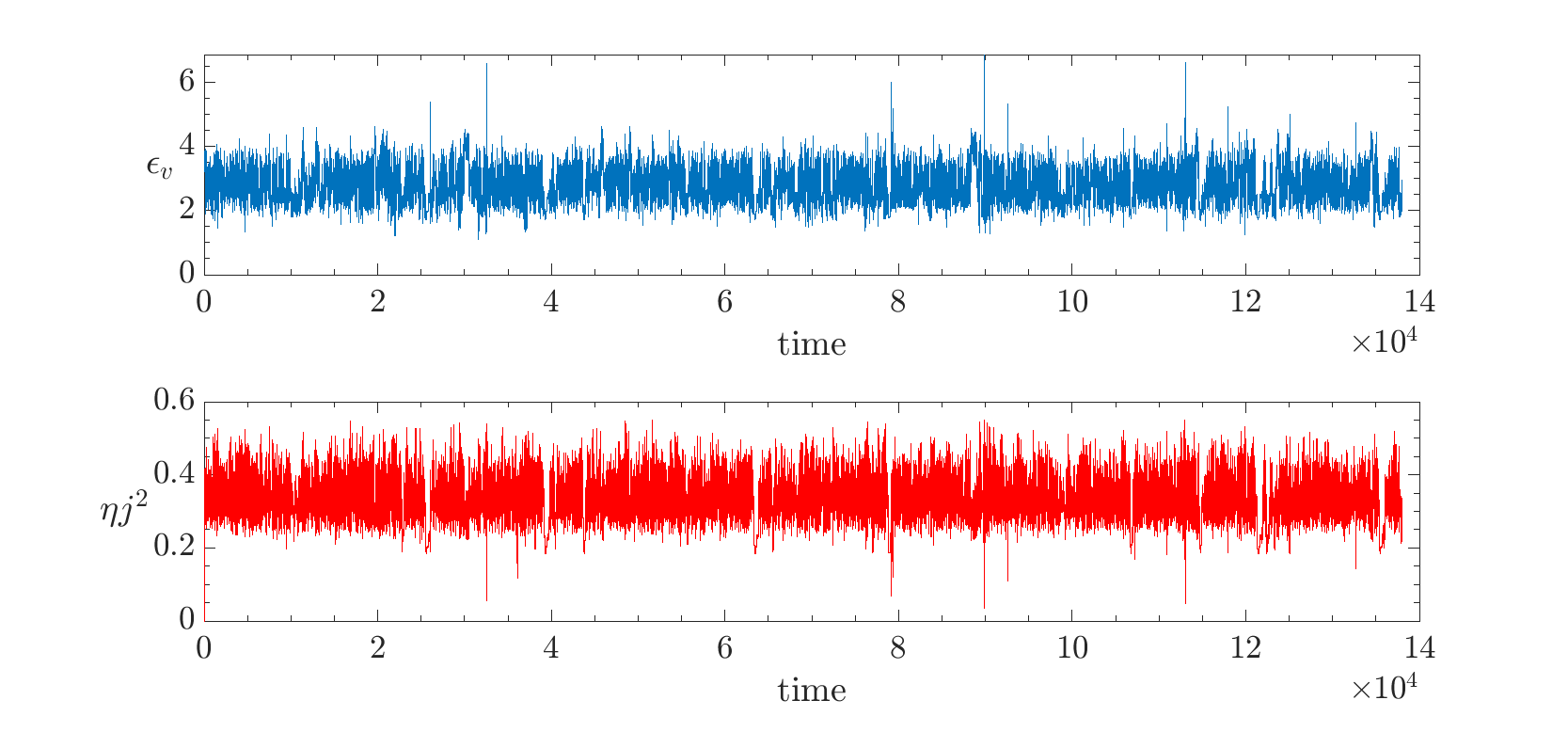} 
  \includegraphics[width=7cm, height=4cm]{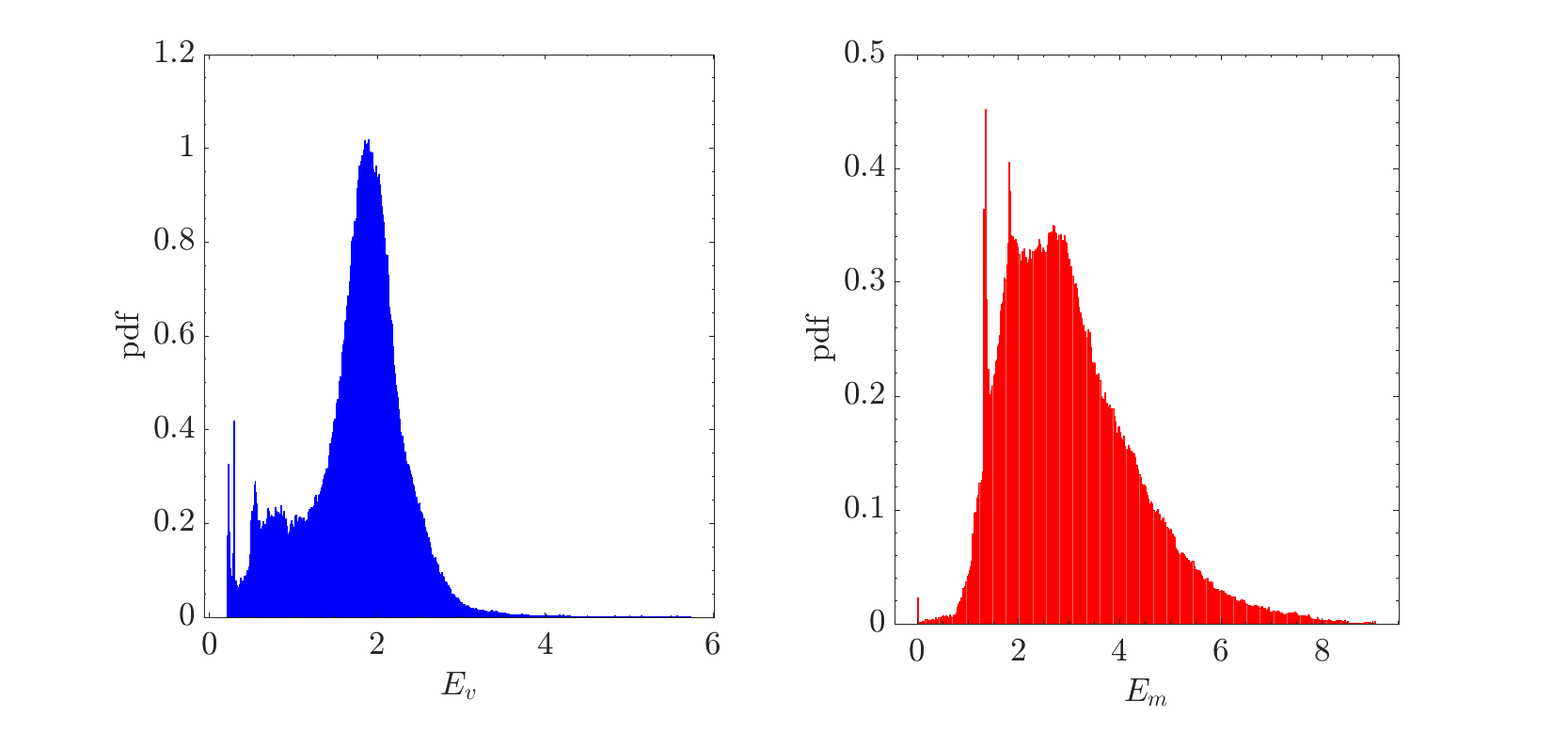}
  \includegraphics[width=7cm, height=4cm]{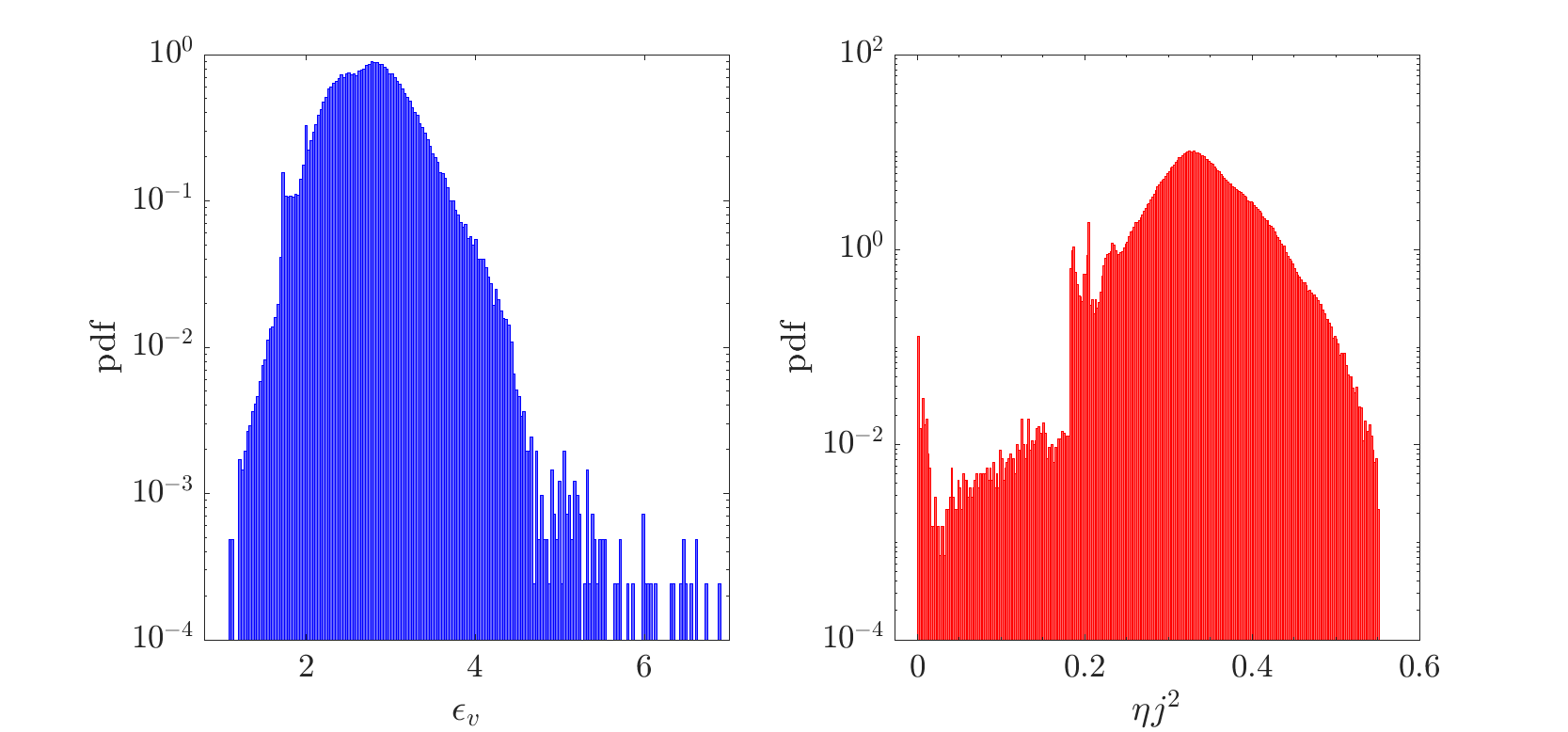}
  \caption{{Similar plots as in Fig. \ref{f:1} but now for the TG3 run.}
}  \label{f:2} 
 \end{figure*}  

%
 \subsection{{Collecting the spatio-temporal data for further analysis}} \label{SS:EXPLAIN}
 {Our methodology is the following.  For each field variable
  at a fixed observation time $T_O$, as  for the vertical component of the velocity $v_z({\bf x}, T_O)$, or the  magnetic energy dissipation $\eta {\bf j}^2({\bf x}, T_O)$,}
   the data is collected roughly every turn-over time $\tau_{nl}$, for in excess of $5 \times 10^4$ outputs,
  { as shown in column $T_{max}$ in Table 1.}
 For each full-cube temporal data  {output at $T_O$}   
 the  {spatially}-dependent three-dimensional field{s} we analyze {are} computed pointwise at each {${\bf x}$} location in the $n_p^3$ data cube.
{PDFs are constructed for these various fields, at that fixed time $T_O$,}  and their second, third and fourth moments are evaluated to {yield} skewness and kurtosis for that time index {$T_O$. These data are then assembled in $K(S)$ plots for the different physical variables of interest.
In Figs. \ref{f:3} and \ref{f:4} below, the time arrow of the data  is indicated by the color  of the points in the scatter plots, with a rainbow  code as given by the color bar at  the left of the plots, with purple at early times and red at late times. We note that,}
having in excess of $5 \times 10^4$ time stamps  {$T_O^{(i)}$} for all runs, the error on the skewness is less than $0.025$, and twice that for the kurtosis. 
 {As expected, these systems achieve statistical equilibria rapidly, in a few $\tau_{nl}$
 (see  Figs. \ref{f:1},\ref{f:2} giving the whole time-span of the runs for the kinetic and magnetic energies).

 Note that this methodology does not correspond to a  spatial analysis of intermittency at a fixed time, such as the first maximum of dissipation, given the moderate spatial resolution of the runs analyzed herein. This  would lead to a study of the localization and intensity of extreme events at that time, but this is not the purpose of this study. \\
Rather, we  instead consider solely here the}
statistical variations over long times of the spatial intermittency of  several field  statistics in a global way.
We do that in terms of overall 
variation of nondimensionalized moments of various functionals of the velocity and magnetic fields as they  evolve in time. With samples taken roughly every turn-over time, the statistical data points can be considered as  independent. This leads to an  investigation of the scatter in values of both $K$ and $S$ for a sample of field functionals, and thus of the  $K(S)$ relationships  that may  emerge for them, with each point on the scatter plots of Figs. \ref{f:3} and \ref{f:4} corresponding to a given $T_O$.
We  finally note that these analyses require substantial storage capabilities even at moderate spatial resolutions, so that the various statistical variables of interest and their moments are better computed {\it a priori}, and stored as the runs progress.
 
 \subsection{{Overall data analysis}}
 
We give for the ABC2 flow (Figure \ref{f:1}) and for the TG3 case (Figure \ref{f:2}) the temporal evolution of the kinetic (top) and magnetic (middle) energy as a function of time 
{on the left, and on the right of their respective total dissipations, $\epsilon_{V,M}$.}
 Note that time is expressed in output counts, with a turn-over time being roughly twice that. 
 At the bottom are given the energy PDFs for the velocity ({blue, leftmost}) and the magnetic field {(red, rightmost plots.})
  In  {all} 
   cases, there are sustained fluctuations in the amplitude of the fields, and in the case of the ABC run, lapses in both kinetic and magnetic energy corresponding to the on-off mechanism. This is directly related to the fact that the PDFs, in that case, have two relative maxima (with one peak close to zero for the magnetic field), whereas in the case of the TG forcing, the structure of the PDFs is 
   {closer to being unimodal. We note that exponential fits in the case of the dissipation fields are plausible, but not yet very strongly marked given the moderate range of Reynolds numbers in these runs.
   }

\begin{figure*}   
  \includegraphics[width=14.0cm, height=7.5cm]{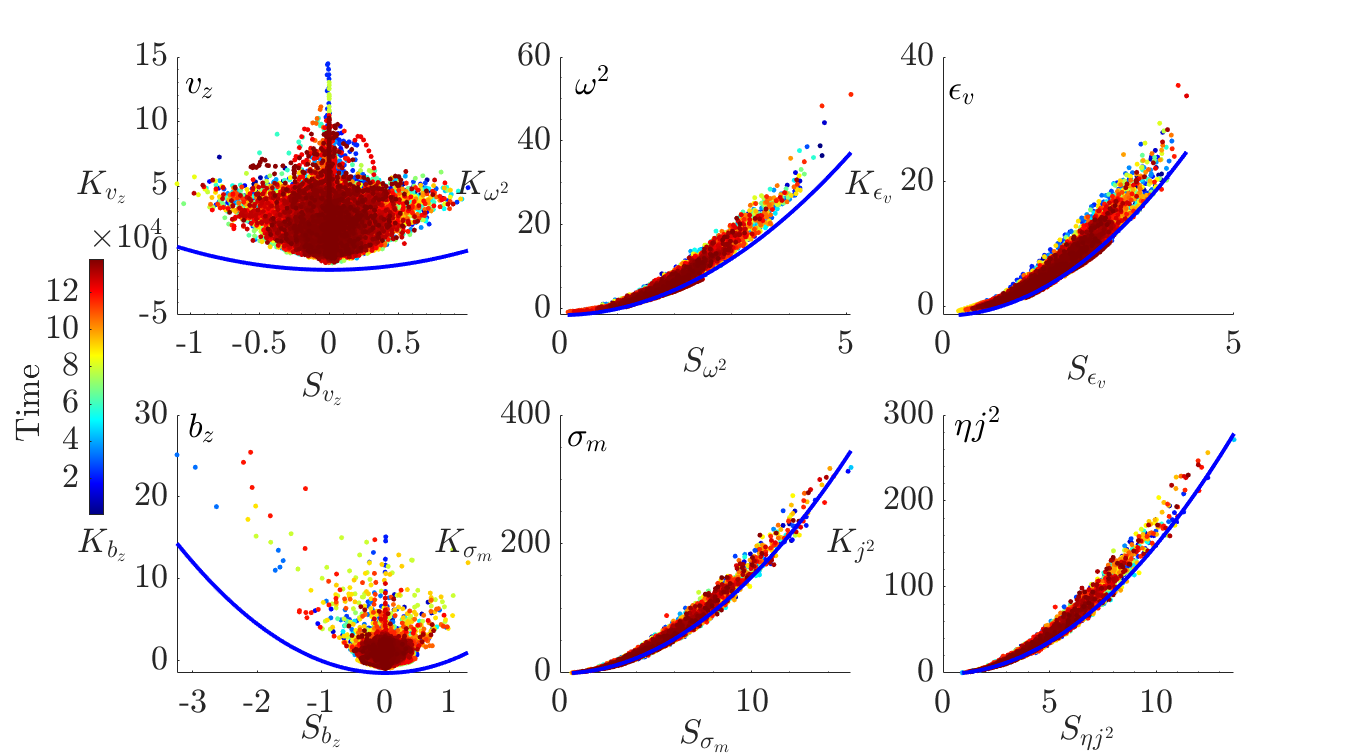}        \hskip0.05truein   
  \caption{
  {\underline{Top:}} For run TG3,   $K(S)$ for the vertical velocity $v_z$ (left),  the square vorticity density  $\vomega^2$ (middle), and the point-wise kinetic energy dissipation $\epsilon_v$ (right).
{\underline{Bottom:}} $K(S)$ for  $b_z$  (left), $\sigma_m$ (middle), and  $\eta {\bf j}^2$  (right). 
The color bar at left indicates the temporal clock in units of turn-over times, with early (late) times in blue (red). The blue lines follow $K(S)=3/2[S^2 - 1]$. 
 }  \label{f:3} \end{figure*}

Fig. \ref{f:3}  gives $K(S)$ for various fields; at top, we display the $K(S)$ relationship for the z-component of the velocity, the square vorticity and the kinetic energy dissipation, namely 
{$K_{v_z}({\bf x}), K_{\omega^2}({\bf x})$ and $K_{\epsilon_v}({\bf x})$.
}
At bottom we plot the equivalent fields for the magnetic induction, namely {$K_{b_z}({\bf x}), K_{\sigma_m}({\bf x})$ and 
$K_{\eta {\bf j}^2}({\bf x})$.
}
 The blue lines correspond to $K=[3/2][S^2-1]$ as mentioned before (see 
\S \ref{SS:MOD}).

We  note the following: whereas for Navier-Stokes turbulence, the three components of the velocity field are Gaussian with the $K(S)$ relation  centered on the $(0,0)$ point,  
 here the peak in values for $K$ at $S\approx 0$ for $v_z$ is up to $K\approx 14$, and rather narrowly centered around $S_{v_z}\approx 0$; high $K$ values are  also present for $b_z$. 
The hydrodynamic case analysed in \citet{pouquet_23a} is computed at comparable $R_\lambda$ and $n_p$ (but not $T_{max}$), and both $S$ and $K$ for  the velocity are  close to $0$. On the other hand, for stratified flows with or without rotation, the vertical component of the velocity, $v_z$, can have high kurtosis and  high values itself; this is associated with the intermittency of dissipation because of the variability of the system dominated by waves {with the sudden development of} small dissipative scales through shear-related instabilities. For this MHD run, the  $x$ and $y$ components of the velocity  behave approximately in the same way as $v_z$ (not shown), with a skewness comparable to that of $v_z$ but smaller kurtosis.

The symmetric and anti-symmetric parts of both the velocity and magnetic field gradient tensors have both high skewness and high kurtosis and for them, a quasi-parabolic fit is appropriate.
 The magnetic dissipation has higher kurtosis and thus is likely significantly more intermittent than its kinetic counterpart, and similarly for $\sigma_m$ and vorticity. On the other hand, both kinetic and magnetic dissipations have lower skewness and kurtosis than for the other part of their gradient tensors, although their statistics overall are similar.
We recall that double exponential (Laplace), or Weibull distributions have small $S$ of either sign and high kurtosis (\cite{bertin_06, biri_15, aschwanden_16}). In that context, we note that it  is shown in \cite{sorriso_18} that a proxy of energy transfer for the solar wind can be defined based on exact laws for MHD corresponding to the conservation of energy and cross-helicity 
$H_C=\left< {\bf v} \cdot {\bf b} \right>$; these proxy fields display high intermittency in {\it Helios 2} (and Ulysses) data, with plausible stretched exponential fits. A final remark is that data points with $[K,S]\approx 0$
must be dominated by random noise at these times; they could correspond to relaxation intervals following  sharp bursts in energy dissipation.

 \begin{figure*}   
  \includegraphics[width=12.9cm, height=7.5cm]{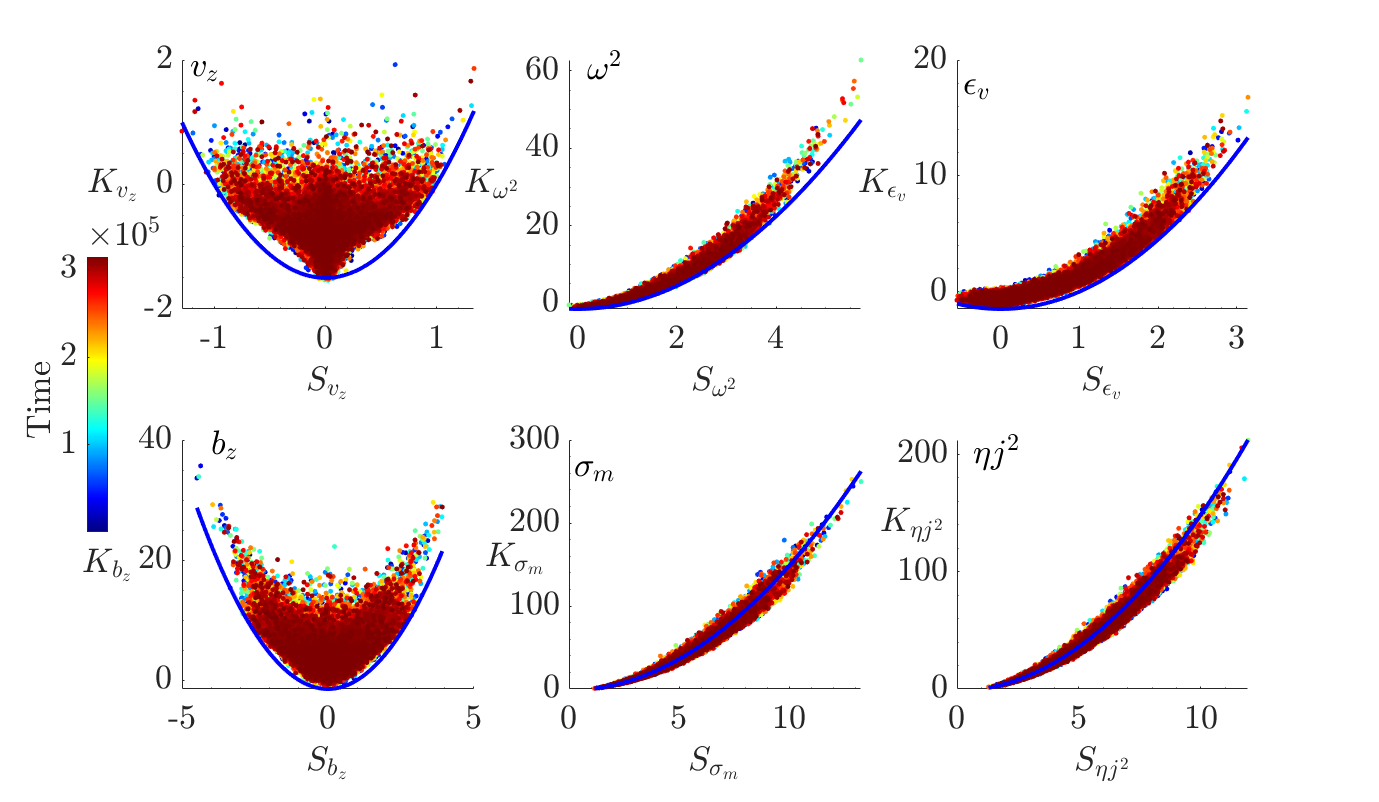}       
 \includegraphics[width=12.0cm, height=3.5cm]{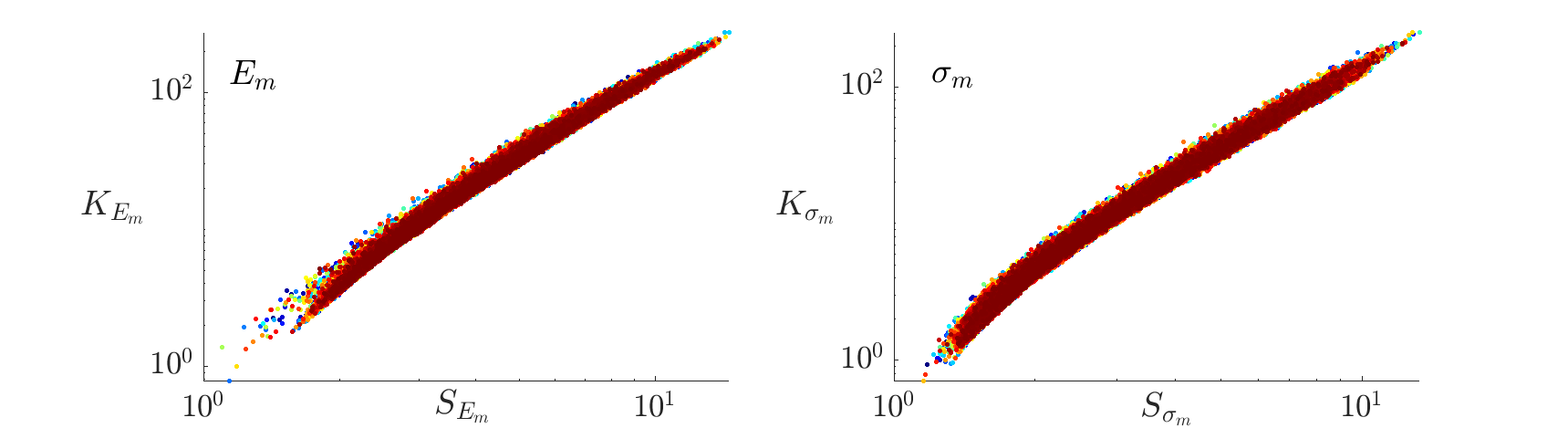}           
  \caption{
The first two rows are the same as Fig. \ref{f:3} for run ABC2 with $R_V\approx 175$ and $T_{max}\approx 3.14 \times 10^5$.
{Bottom:   $K(S)$ is plotted for the magnetic energy $E_m$ (left) and for 
the symmetric part of the magnetic field gradient tensor
$\sigma_m$ (right),  using now log-log coordinates.
 Approximate fits give for both  $\alpha \approx 2.2, \kappa \approx 1.0$.
}
 }  \label{f:4} \end{figure*}   

We now check whether this behavior is observed as well for another type of forcing. In Fig. \ref{f:4} are plotted the $K(S)$ relationships at the top for $v_z$ (left), $\vomega^2$ (middle) and $\epsilon_v$ (right) and (middle row),
 $b_z$ (left), $\sigma_m$ (middle) and $\eta {\bf j}^2$ (right), as in Fig. \ref{f:3} but now for run ABC2 with a fully helical ABC forcing.  
Values for $(S,K)$ for both runs are comparable except for the vertical component of the velocity due to its specific structure. 
{We added in the last row of Fig. \ref{f:4} the $K(S)$ plots for the magnetic energy (left) and
$\sigma_m$ (right) for the same run, and using now log-log coordinates.
Power laws are conspicuous above some threshold  in skewness, with 
$E_m \approx 0.96S^{2.20}, \ \sigma_m \approx  0.96S^{2.25},\ \eta j^2\approx 1.05 S^{2.17}$.
}
 
Finally, due to the strong symmetries of the initial conditions and forcing of the dynamos analyzed in this paper, one could wonder whether the addition of a small noise would change the results. On the other hand, in view of the length of the computations, well beyond what can be estimated for a reasonable Lyapounov time of separation of trajectories, it is unlikely that the overall results, and in particular the quasi-parabolic law for dissipation, would be altered. Indeed, one can find estimates of the first Lyapounov exponent $\lambda_1$, in the ABC dynamo for example, with $\lambda_1\sim 0.073$ for a run with $R_V\approx 60, P_M=4$, comparable to what we have here (\cite{zienicke_98, alexakis_08}), meaning that after  roughly $14 \tau_{nl}$, the initial conditions have been forgotten.

\section{Kurtosis-Skewness law as given by classical intermittency models} \label{S:KS}
We can in fact  compute the scaling exponent $\alpha_f$ in the relationship $K(S)\sim S^{\alpha_f}$, assuming  the usual formulation for the structure functions of order $p$ for a scalar field $f$, namely the field differences over a distance $r$, $\left<\delta f(r)^p \right> \sim r^{\zeta_p}$; one  obtains  for 
$\alpha_f$: 
\be
\alpha_f=\frac{\zeta_4 -2\zeta_2}{\zeta_3 -\frac{3}{2}\zeta_2} \ . \ee
Within the framework of the standard multi-fractal \citet{she_94} (SL)  
{ model for fluids (slf)}, and generalized for MHD in \citet{grauer_94, politano_95pr} {(slm)}, one has:
\bea
 \zeta_p^{slf}&=& \frac{p}{9} + 2 \left[ 1 - \left( \frac{2}{3} \right)^{p/3} \right] \  ; \  \  \alpha_{slf} = \frac{2 [1 - 2(2/3)^{2/3} +  (2/3)^{4/3}]}{7/3 - 3 (2/3)^{2/3}} \ \approx \ 2.56  \ ; \label{SLF} \\
 \zeta_p^{slm}&=& \frac{p}{8} + 1 -  \left( \frac{1}{2} \right)^{p/4} \ \ \ \ \ \ \ ; \ \ \alpha_{slm} = \frac{3 - 4 (1/2)^{1/2}}{1+ 2 (1/2)^{3/4} - 3 (1/2)^{1/2}}  \ \approx \ 2.53 \ .     
 \label{SLM} \eea 

In building these SL models for fluid turbulence ({\it slf}) and MHD ({\it slm}), an assumption is made that a hierarchy of flux structures exists compatible with a Kolmogorov transfer timescale and with vortex filaments (or in MHD, an isotropic Iroshnikov-Kraichnan wave-eddy interaction and current sheets). This leads to a specific non-linear relation in $p$ for the $\zeta_p$s.  Note that a  parabolic scaling, $\alpha=2$, is obtained when considering  the generalized versions of these log-Poisson models  -- derived in \citet{politano_95pr} both for fluids and for MHD -- as the intermittency becomes maximal with extreme flux structures 
whose geometrical signature disappears in the expression of $\alpha$ (see  \citet{pouquet_24}).

\begin{figure*}  
  \includegraphics[width=7cm, height=5cm]{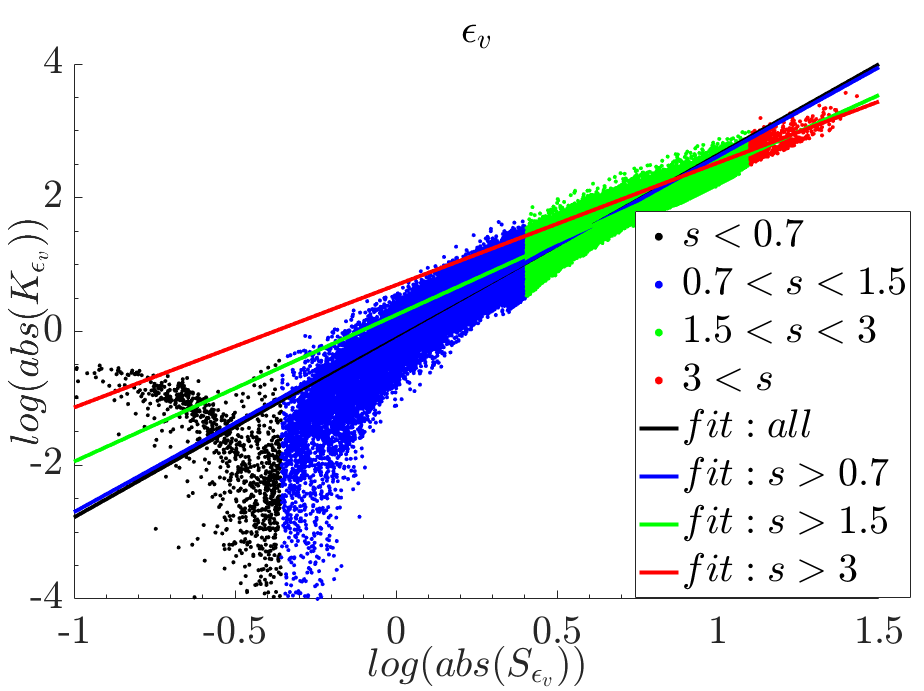} 
  \includegraphics[width=7cm, height=5cm]{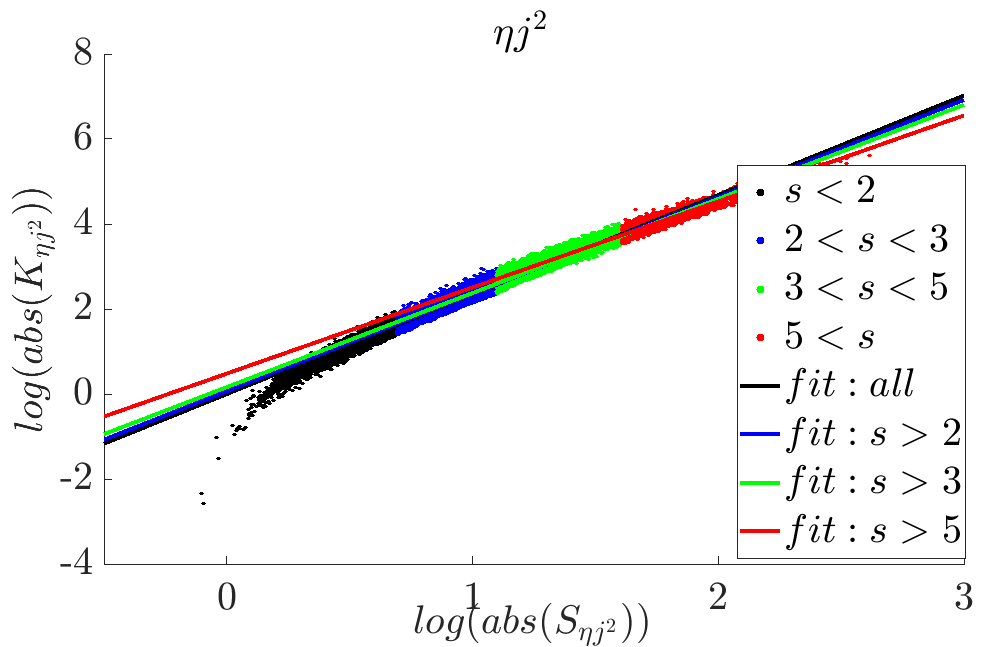}
  \includegraphics[width=7cm, height=5cm]{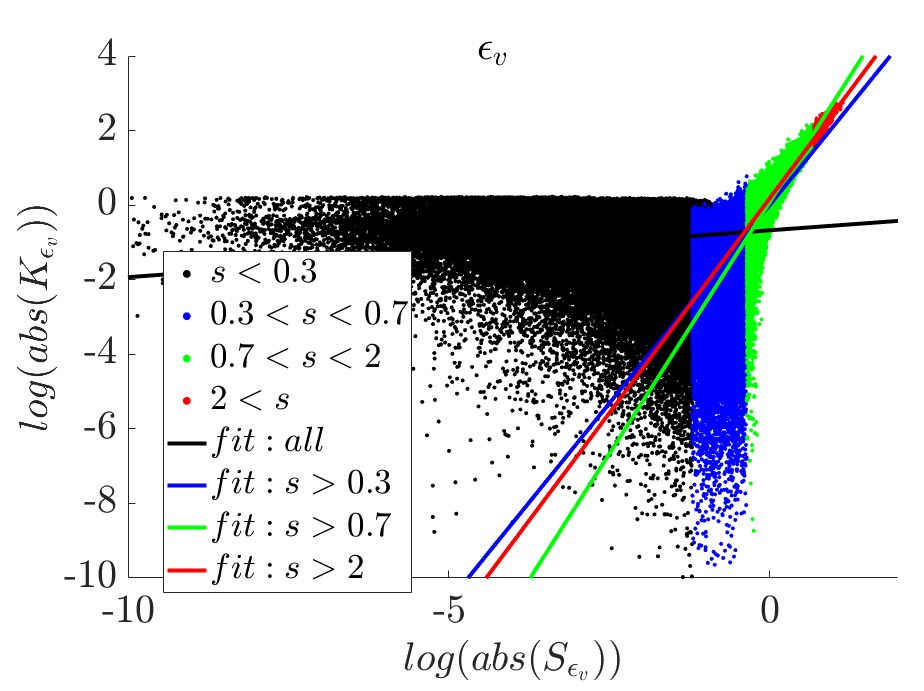}  
  \includegraphics[width=7cm, height=5cm]{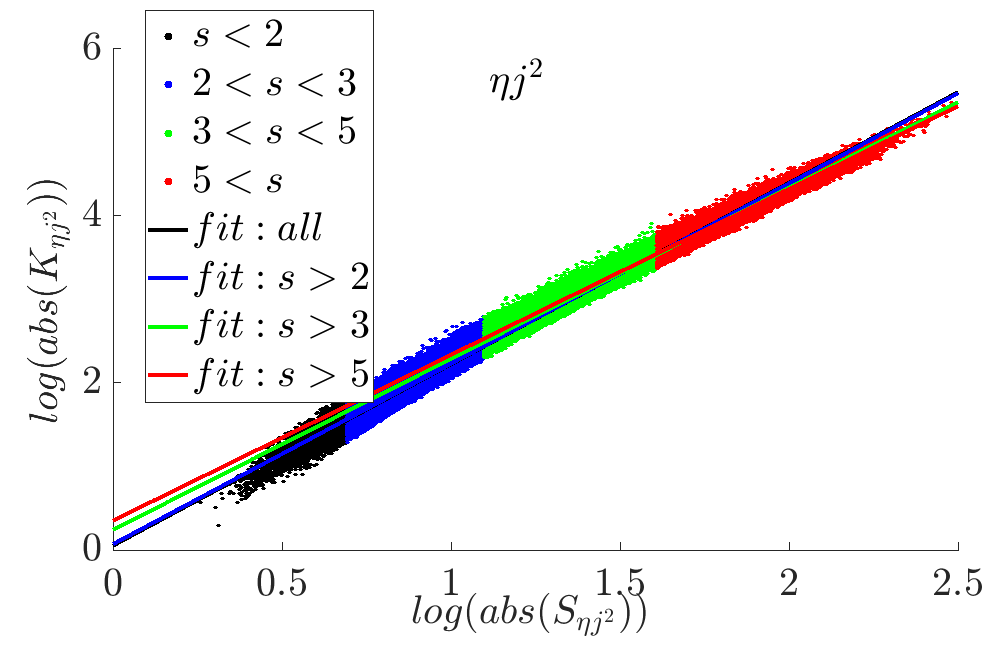}
  \caption{ 
Log-log plot for  $|K|(|S|) = \kappa |S|^{\alpha}$,  runs TG3 (top) and  ABC2 (bottom), for kinetic (left) and magnetic (right) dissipation.  
Thresholds in $S$ are displayed in different colors, 
{and similarly for the power-law fits, as indicated in each insert.}
  }  
  \label{f:5} 
 \end{figure*}  

\begin{figure*}   
  \includegraphics[width=12cm, height=4cm]{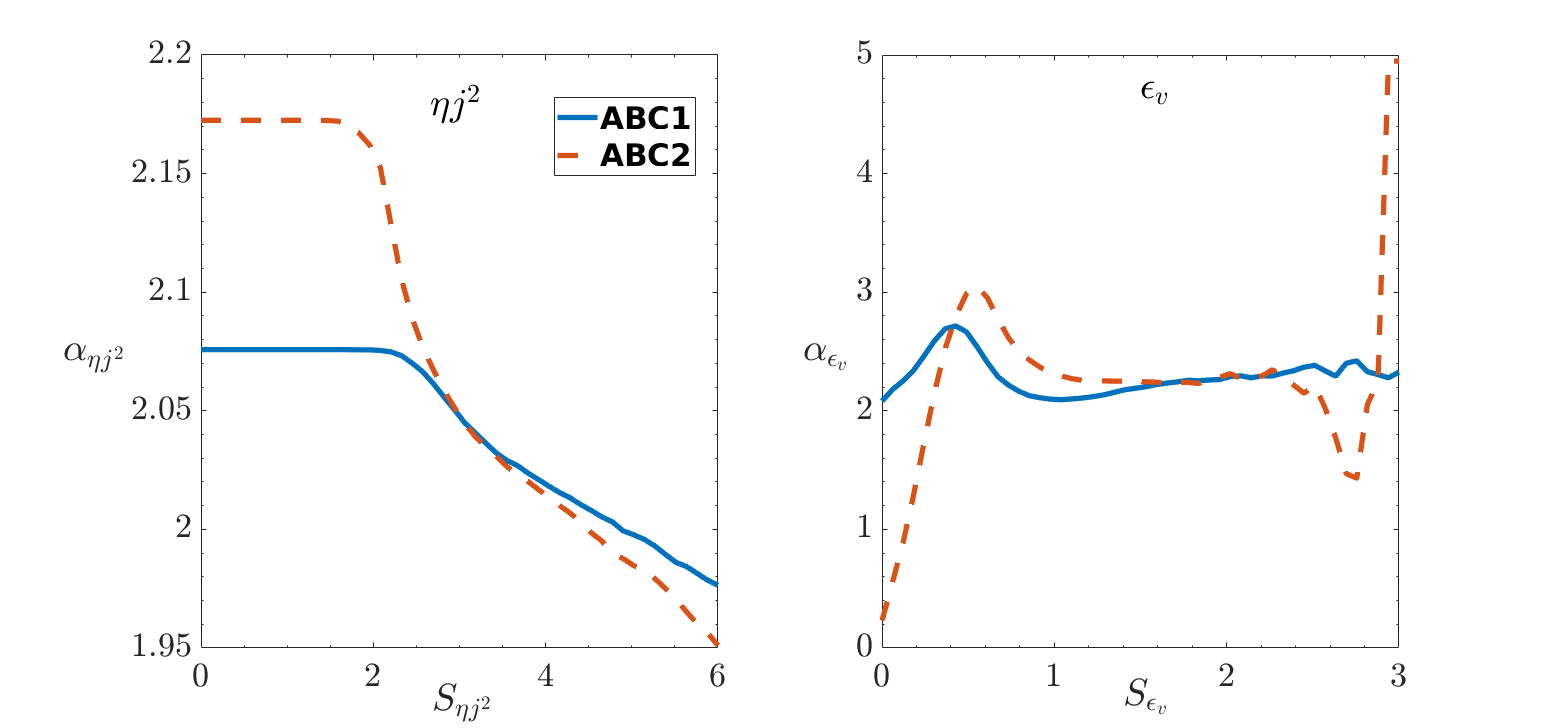}
  \includegraphics[width=12cm, height=4cm]{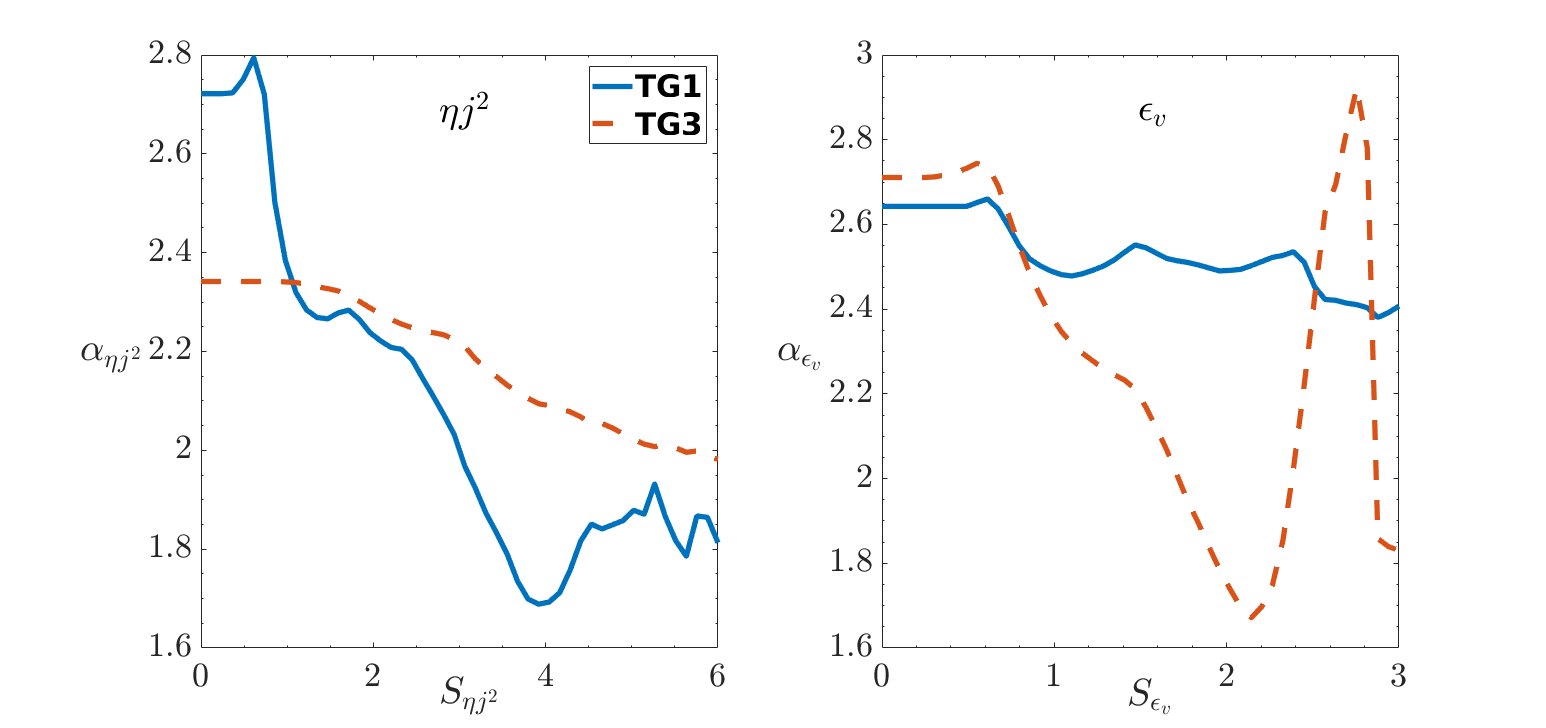}
  \caption{
$|K(S)|\sim \kappa|S|^\alpha$ fit:  variations with  filter threshold  in $S$ of  $\alpha$ for $\eta {\bf j}^2$ (left) and $\epsilon_V$ (right) for the  ABC  (top) and TG  (bottom) flows; runs are differentiated by their colored lines.
  }  
  \label{f:6} 
 \end{figure*}

In this context, we compute  power law fits, $|K| = \kappa |S|^\alpha$, as displayed in Figs. \ref{f:5} and \ref{f:6}) for the  TG3  and ABC2  runs for the kinetic  and magnetic  dissipations, 
and for both full and thresholded data.  There are clear variations of $\alpha$ with the strength of the intermittency as evaluated through the level of the skewness. 
Specifically, the chosen thresholds  are $S<2$ (black), $2\le S <3$ (blue), $3\le S < 5$ (green) and $S\ge 5$ (red). Power-law fits in these  intervals are given using the same colors. The high $S, K$ values reached here are related to the very long integration times allowing for a thorough exploration  of  configuration space. We note that the velocity field has a broad range with non-intermittent values (black dots) in the sense that both $S$ and $K$ are quite low; it also undergoes a change of sign of the skewness at low values for TG3. We also observe that  $S$ and $K$ can be substantially higher for $\eta j^2$ than for the point-wise kinetic energy dissipation, and with less scatter for $K$ at a given $S$. This difference might be related to the dynamo 
that controls the behavior of the magnetic field and its dissipation, whereas the velocity, at these resolutions, still feels the effect of the forcing, but the fit that includes these low-$S$ points does not represent the behavior of the more intermittent data.

Skewness and excess Kurtosis can be negative, which is why absolute values are used to visualize the realizations. It is important to mention that Fisher's definition is applied here (where a normal distribution corresponds to zero), then negative values (above $-3$) are common. For magnetic field quantities, the percentage of negative skewness (S) or kurtosis (K) is nearly zero. In contrast, the skewness of the velocity dissipation rate $\epsilon_v$ can reach up to 2\% negative values across all realizations (thus represent only few black points in Fig \ref{f:5}). Only in the ABC2 case, the proportion of negative and positive kurtosis values of the velocity dissipation becomes equivalent.
    
We give some more quantitative information in Fig. \ref{f:6}, looking at variations with threshold in skewness of the 
{power-law index $\alpha$  for the dissipation fields, namely $\eta j^2$ (left) and $\epsilon_v$ (right) for the two ABC  runs (top) and the two TG runs (bottom); these runs are identified in the inserts by line and color.
}

We note first that the run TG3 has a higher $R_M$, and the two ABC runs have higher $R_\lambda$.
{We also note that the overall ranges of variations for the power-law index is lesser for the TG runs than for the ABC runs, probably due to the fact that the TG run is more developed, and the effect of the forcing is lesser.}
The power-law index  for $\epsilon_v$ for run ABC2 has a substantially larger range of variation than for $ \eta j^2$, with  $0^+ <  \alpha_{\epsilon_v} \lesssim 5.0$ overall, {\it versus}
  $1.95 <  \alpha_{j^2} \lesssim 2.17$. 
{In general, rather abrupt changes in the values of  $\alpha$ (and $\kappa$, not shown) 
 occur for $S\gtrapprox 0.5$, {\it i.e.} when turbulent motions develop locally.}
For the current, $\alpha_{\eta j^2}$ decreases systematically towards $\alpha_{\eta j^2}\approx 2$ when the threshold in 
{$S_{\eta j^2}$}
 is increased. 
{We also note that} 
 we see a systematic decreasing trend in $\alpha$ towards a value of $2$ or slightly lower, a value that can be recovered with the extension of the SL model to more varied dissipative structures (\cite{pouquet_24}). 
Finally, the constant $\kappa$ (not shown) is of order unity in all cases, and slightly increases with threshold as long as enough data is available. We also note its quasi-constancy at lower $S$ values.

 A common feature of all these plots is that 
 {there are notable}
  variations with threshold in $S$, 
  {starting rather abruptly in the case of the ABC flows.}
   This could indicate an effect on the velocity of the influence of the forcing at these moderate Reynolds numbers and that 
    the velocity is not necessarily changed by the magnetic field which remains somewhat weak (see Figs. \ref{f:1} and \ref{f:2} and Table 1). Indeed, with the ratio $r=R_M/R_M^C\approx 1^+ $ for all runs except run TG3 (see Table 1), we are still rather close to the threshold of the onset of the dynamo.
 \section{Discussion, conclusion and perspectives} \label{S:C}
 %

We have found in this paper that, 
{rather close to the threshold of} 
 the fast dynamo regime,  a now classical quasi-parabolic behavior between kurtosis and skewness  
is present for kinetic and magnetic variables.
The numerical data we analyzed  represent but one  aspect of the study of intermittency in the dynamo regime, and many questions remain. One issue concerns the effect a fully turbulent velocity will have on $K(S)$ scaling  and the turbulence of the magnetic field itself. For example, using  wavelets, \cite{camporeale_18p} could estimate in high-resolution 2D DNS of Hall-MHD that only 25\% of the volume supports 50\% of the energy transfer, giving thus a quantitative estimate of the intermittency of  energy dissipation; we note that, for stratified fluids, this proportion can go as low as 11\% of the kinetic energy dissipation for high kurtosis of the vertical velocity (\cite{marino_22}). It will be of interest to examine as well  these statistics in the case of fast dynamos at higher Reynolds numbers.
  

Another question is   whether the intermittency of the early dissipation range dominates the statistics, at least at moderate Reynolds numbers. Indeed, one could argue that the intermittency of the dissipation is mostly located in the beginning of the range, due to the ensuing fast decay. In \citet{hu-hh_23}, a near-dissipation range intermittency is examined using Solar Wind data obtained with the Parker Solar Probe. The authors conclude that they find evidence for log-Poisson scaling as modeled for MHD in \cite{grauer_94, politano_95pr}, and that such structures are also almost entirely responsible for the intermittency anisotropy (see also \cite{bian_24}). This may be consistent with stating, as   developed already in  \citet{kraichnan_67b}, that most of the intermittency is in fact at the beginning of the dissipative range.

Another issue is whether or not there is a dynamical consequence of $K(S)$ being close to its Cauchy-Schwarz limit. These inequalities linking $K$ and $S$ can be viewed as a limitation both on skewness (which has to be smaller than some value) and kurtosis, which cannot be too small. It  shows that non-Gaussianity and intermittency are unavoidable (except for the trivial $(K=0,S=0)$ solution), but also that intermittency is limited in the sense that $K$ and $S$ are not independent, and at least for the NS case, the skewness is constrained by the exact law stemming from energy conservation (\cite{K41c}), the laws in MHD involving cross-correlations (see for a recent review \cite{marino_23r}). Furthermore, as noted by several authors,  $K(S)$ laws may put some limitation on the type of PDFs that a particular intermittent field follows. Similarly, some of these $K(S)$ relationships may contribute   insight as to the relevance of  Large Eddy Simulation  (LES) parametrizations by providing  constraints on the flow characteristics.

 The role of anisotropy in interpreting the dynamics of turbulent flows  is complex, including at larger $R_V$ such as that encountered in the atmosphere (\cite{lovejoy_01}).  For example, it is shown in \cite{galtier_23} that it affects in different ways the amplitude of the energy distribution and the spectral indices, so more work will be needed in that direction as well; we already know that, for anisotropic fluid flows in the presence of stratification, the skewness and kurtosis of velocity components can be direction-dependent (\cite{bos_07}; see also \cite{homann_14} for the fast dynamo), and that anisotropic scaling laws can be developed phenomenologically and found observationally (\cite{nazarenko_11, bian_24}). 

When the velocity field is chaotic but not yet fully turbulent, and close to threshold for dynamo action,  \cite{sweet_01}  identified a temporal bursty `on-off" behavior of the dynamo-generated magnetic field which grows on average linearly with the control parameter, {\it i.e.} the distance in $R_M$ from the threshold (see also \cite{ponty_07p, alexakis_08}, but these authors did not give indications on the behavior of the first few moments of the growing field).
In \cite{alexakis_08},  the Lorentz force feed-back on the flow is studied in detail with DNS ran for up to $10^5 \tau_{nl}$  and for various 
 $P_M$. They  find that the Lorentz force strongly modifies the temporal evolution of the growing field through a control of the noise.
We already know that the noise can alter 
 the coefficients in a parabolic relation (\cite{theodorsen_17, losada_23}), so we might be able to observe  a change of scaling once we enter a turbulent saturation regime for the dynamo at higher Reynolds numbers, as we did for stratified flows (\cite{pouquet_23a}). This is left for future work. 
\\

{\it Acknowledgements:} Yannick Ponty thanks A. Miniussi  for computing design assistance on the CUBBY code. The authors are grateful to the OPAL infrastructure from Université Côte d’Azur, the Université Côte d’Azur’s Center for High-Performance Computing, and to the national French computer facilities (GENCI) for providing resources and support. Annick Pouquet is thankful to Bob Ergun for his encouragement.

\medskip
{\it Declaration of interests:}
The authors report no conflict of interest.

 \bibliographystyle{jpp}  
 \bibliography{newmac_040125} 

\begin{thebibliography}{60}
\expandafter\ifx\csname natexlab\endcsname\relax\def\natexlab#1{#1}\fi
\def\au#1{#1} \def\ed#1{#1} \def\yr#1{#1}\def\at#1{#1}\def\jt#1{\textit{#1}}
  \def\bt#1{#1}\def\bvol#1{\textbf{#1}} \def\vol#1{#1} \def\pg#1{#1}
  \def\publ#1{#1}\def\arxiv#1{#1}\def\org#1{#1}\def\st#1{\textit{#1}}

\bibitem[Adhikari {\em et~al.\/}(2020)Adhikari, Shay, Parashar, Pyakurel,
  Matthaeus, Godzieba, Stawarz, Eastwood \& Dahlin]{adhikari_20}
{\sc \au{Adhikari, S.}, \au{Shay, M.~A.}, \au{Parashar, T.~N.}, \au{Pyakurel,
  P.~S.}, \au{Matthaeus, W.}, \au{Godzieba, D.}, \au{Stawarz, J.},
  \au{Eastwood, J.~P.} \& \au{Dahlin, J.~T.}} \yr{2020}  \at{Reconnection from
  a turbulence perspective}.  \jt{Phys. Plasmas}  \bvol{27},  \pg{042305}.

\bibitem[Alexakis \& Ponty(2008)]{alexakis_08}
{\sc \au{Alexakis, A.} \& \au{Ponty, Y.}} \yr{2008}  \at{Effect of the
  {L}orentz force on on-off dynamo intermittency}.  \jt{Phys. Rev. E}
  \bvol{77},  \pg{056308}.

\bibitem[Aschwanden {\em et~al.\/}(2016)Aschwanden, Crosby, Dimitropoulou,
  Georgoulis, Hergarten, McAteer, Milovanov, Mineshige, Morales, Nishizuka,
  Pruessner, Sanchez, Sharma, Strugarek \& Uritsky]{aschwanden_16}
{\sc \au{Aschwanden, M.~J.}, \au{Crosby, N.~B.}, \au{Dimitropoulou, M.},
  \au{Georgoulis, M.~K.}, \au{Hergarten, S.}, \au{McAteer, J.}, \au{Milovanov,
  A.~V.}, \au{Mineshige, S.}, \au{Morales, L.}, \au{Nishizuka, N.},
  \au{Pruessner, G.}, \au{Sanchez, R.}, \au{Sharma, A.~S.}, \au{Strugarek, A.}
  \& \au{Uritsky, V.}} \yr{2016}  \at{25 years of self-organized criticality:
  Solar and astrophysics}.  \jt{Space Sci. Rev.}  \bvol{198},  \pg{47--166}.

\bibitem[Bertin \& Clusel(2006)]{bertin_06}
{\sc \au{Bertin, E.} \& \au{Clusel, M.}} \yr{2006}  \at{Generalized extreme
  value statistics and sum of correlated variables}.  \jt{J. Phys. A}
  \bvol{39},  \pg{7607--7619}.

\bibitem[Bian \& Li(2024)]{bian_24}
{\sc \au{Bian, N.~H.} \& \au{Li, G.}} \yr{2024}  \at{Lagrangian perspectives on
  the small-scale structure of {A}lfvénic turbulence and stochastic models for
  the dispersion of fluid particles and magnetic field lines in the solar
  wind}.  \jt{APJ Supp.}  \bvol{273},  \pg{15}.

\bibitem[Biri {\em et~al.\/}(2015)Biri, Scharffenberg \& Stammer]{biri_15}
{\sc \au{Biri, S.}, \au{Scharffenberg, M.~G.} \& \au{Stammer, D.}} \yr{2015}
  \at{A probabilistic description of the mesoscale eddy field in the ocean}.
  \jt{J. Geophys. Res.}  \bvol{120},  \pg{4778--4802}.

\bibitem[Bos {\em et~al.\/}(2007)Bos, Liechtenstein \& Schneider]{bos_07}
{\sc \au{Bos, W.}, \au{Liechtenstein, L.} \& \au{Schneider, K.}} \yr{2007}
  \at{Small-scale intermittency in anisotropic turbulence}.  \jt{Phys. Rev.}
  \bvol{67},  \pg{046310}.

\bibitem[Bradshaw {\em et~al.\/}(2019)Bradshaw, Farhat \& Grujic]{bradshaw_19}
{\sc \au{Bradshaw, Z.}, \au{Farhat, A.} \& \au{Grujic, Z.}} \yr{2019}  \at{An
  algebraic reduction of the scaling gap in the navier-stokes regularity
  problem}.  \jt{Arch. Rational Mech. Anal.}  \bvol{231},  \pg{1963--2005}.

\bibitem[Brandenburg \& Subramanian(2005)]{brandenburg_05}
{\sc \au{Brandenburg, A.} \& \au{Subramanian, K.}} \yr{2005}  \at{Astrophysical
  magnetic fields and nonlinear dynamo theory}.  \jt{Phys. Rep.}  \bvol{{\bf
  417}}~(1).

\bibitem[Buaria {\em et~al.\/}(2022)Buaria, Pumir \& Bodenschatz]{buaria_22p}
{\sc \au{Buaria, D.}, \au{Pumir, A.} \& \au{Bodenschatz, E.}} \yr{2022}
  \at{Generation of intense dissipation in high reynolds number turbulence}.
  \jt{Phil. Trans. A}  \bvol{380},  \pg{20210088}.

\bibitem[Camporeale {\em et~al.\/}(2018)Camporeale, Sorriso-Valvo, Califano \&
  Retin\'o]{camporeale_18p}
{\sc \au{Camporeale, E.}, \au{Sorriso-Valvo, L.}, \au{Califano, F.} \&
  \au{Retin\'o, A.}} \yr{2018}  \at{Coherent structures and spectral energy
  transfer in turbulent plasma: A space-filter approach}.  \jt{Phys. Rev.
  Lett.}  \bvol{120}~(125101).

\bibitem[Chen(2016)]{chen_16}
{\sc \au{Chen, C.}} \yr{2016}  \at{Recent progress in astrophysical plasma
  turbulence from solar wind observations}.  \jt{J. Plasma Phys.}  \bvol{82},
  \pg{535820602}.

\bibitem[Ergun {\em et~al.\/}(2020)Ergun, Ahmadi, Kromyda, Schwartz, Chasapis,
  Hoilijoki, Wilder, Stawarz, Goodrich, Turner, Cohen, Bingham, Holmes,
  Nakamura, Pucci, Torbert, Burch, Lindqvist, Strangeway, {Le Conte}l \&
  Giles]{ergun_20}
{\sc \au{Ergun, R.~E.}, \au{Ahmadi, N.}, \au{Kromyda, L.}, \au{Schwartz,
  S.~J.}, \au{Chasapis, A.}, \au{Hoilijoki, S.}, \au{Wilder, F.~D.},
  \au{Stawarz, J.~E.}, \au{Goodrich, K.~A.}, \au{Turner, D.~L.}, \au{Cohen,
  I.~J.}, \au{Bingham, S.~T.}, \au{Holmes, J.~C.}, \au{Nakamura, R.},
  \au{Pucci, F.}, \au{Torbert, R.~B.}, \au{Burch, J.~L.}, \au{Lindqvist,
  P.-A.}, \au{Strangeway, R.~J.}, \au{{Le Conte}l, O.} \& \au{Giles, B.~L.}}
  \yr{2020}  \at{Observations of particle acceleration in magnetic
  reconnection–driven turbulence}.  \jt{Astrophys. J.}  \bvol{898},
  \pg{154}.

\bibitem[Furno {\em et~al.\/}(2015)Furno, Avino, Bovet, Diallo, Fasoli,
  Gustafson, Iraji, Labit, Loizu, M\"uller, Plyushchev, Podesta, Poli, Ricci \&
  Theiler]{furno_15}
{\sc \au{Furno, I.}, \au{Avino, F.}, \au{Bovet, A.}, \au{Diallo, A.},
  \au{Fasoli, A.}, \au{Gustafson, K.}, \au{Iraji, D.}, \au{Labit, B.},
  \au{Loizu, J.}, \au{M\"uller, S.~H.}, \au{Plyushchev, G.}, \au{Podesta, M.},
  \au{Poli, F.~M.}, \au{Ricci, P.} \& \au{Theiler, C.}} \yr{2015}  \at{Plasma
  turbulence, suprathermal ion dynamics and code validation on the basic plasma
  physics device torpex}.  \jt{J. Plasma Phys.}  \bvol{81},
  \pg{10.1017/S0022377815000161}.

\bibitem[Galtier(2023)]{galtier_23}
{\sc \au{Galtier, S.}} \yr{2023}  \at{Fast magneto-acoustic wave turbulence and
  the {I}roshnikov-{K}raichnan spectrum}.  \jt{J. Plasma Phys.}  \bvol{89},
  \pg{905890205}.

\bibitem[Garcia(2012)]{garcia_12}
{\sc \au{Garcia, O.~E.}} \yr{2012}  \at{Stochastic modeling of intermittent
  scrape-off layer plasma fluctuations}.  \jt{Phys. Rev. Lett.}  \bvol{108},
  \pg{265001}.

\bibitem[Grauer {\em et~al.\/}(1994)Grauer, Krug \& Mariani]{grauer_94}
{\sc \au{Grauer, R.}, \au{Krug, J.} \& \au{Mariani, C.}} \yr{1994}  \at{Scaling
  of high-order structure functions in magnetohydrodynamic turbulence}.
  \jt{Phys. Lett. A}  \bvol{195},  \pg{335--338}.

\bibitem[Guszejnov {\em et~al.\/}(2013)Guszejnov, Lazanyi, Bencze \&
  Zoletnik]{guszejnov_13}
{\sc \au{Guszejnov, D.}, \au{Lazanyi, N.}, \au{Bencze, A.} \& \au{Zoletnik,
  S.}} \yr{2013}  \at{On the effect of intermittency of turbulence on the
  parabolic relation between skewness and kurtosis in magnetized plasmas}.
  \jt{Phys. Plasmas}  \bvol{20},  \pg{112305}.

\bibitem[van Haren {\em et~al.\/}(2016)van Haren, Cimatoribus, Cyr \&
  Gostiaux]{vanharen_16g}
{\sc \au{van Haren, H.}, \au{Cimatoribus, A.~A.}, \au{Cyr, F.} \& \au{Gostiaux,
  L.}} \yr{2016}  \at{Insights from a 3-{D} temperature sensors mooring on
  stratified ocean turbulence}.  \jt{Geophys. Res. Lett.}
  \bvol{43}~(DOI:10.1002/2016GL068032),  \pg{1--7}.

\bibitem[Hasselmann(1962)]{hasselmann_62-1}
{\sc \au{Hasselmann, K.}} \yr{1962}  \at{On the non-linear energy transfer in a
  gravity-wave spectrum. {P}art 1. {G}eneral theory}.  \jt{J. Fluid Mech.}
  \bvol{12},  \pg{481--500}.

\bibitem[Homann {\em et~al.\/}(2014)Homann, Ponty, Krstulovic \&
  Grauer]{homann_14}
{\sc \au{Homann, H.}, \au{Ponty, Y.}, \au{Krstulovic, G.} \& \au{Grauer, R.}}
  \yr{2014}  \at{Structures and {L}agrangian statistics of the {T}aylor-{G}reen
  dynamo}.  \jt{New J. Phys.}  \bvol{16},  \pg{075014}.

\bibitem[Hughes {\em et~al.\/}(2010)Hughes, Thompson \& Wilson]{hughes_10}
{\sc \au{Hughes, C.~W.}, \au{Thompson, A.~F.} \& \au{Wilson, C.}} \yr{2010}
  \at{Identification of jets and mixing barriers from sea level and vorticity
  measurements using simple statistics}.  \jt{Ocean Mod.}  \bvol{32},
  \pg{44--57}.

\bibitem[Imazio {\em et~al.\/}(2023)Imazio, Mininni, Godoy, Rivaben \&
  Dörnbrack]{rodriguez_23}
{\sc \au{Imazio, P.~R.}, \au{Mininni, P.~D.}, \au{Godoy, A.}, \au{Rivaben, N.}
  \& \au{Dörnbrack, A.}} \yr{2023}  \at{Not all clear air turbulence is
  {K}olmogorov -- {T}he fine-scale nature of atmospheric turbulence}.  \jt{JGR
  Atmospheres}  \bvol{128},  \pg{e2022JD037491}.

\bibitem[Kolmogorov(1941)]{K41c}
{\sc \au{Kolmogorov, A.~N.}} \yr{1941}  \at{Dissipation of energy in locally
  isotropic turbulence \ [{E}nglish translation in {P}roc. {R}oy. {S}oc.
  {L}ondon {A} 434 (1991) 15-17.]}.  \jt{Dokl. Akad. Nauk SSSR}  \bvol{32},
  \pg{19--21}.

\bibitem[Kraichnan(1967)]{kraichnan_67b}
{\sc \au{Kraichnan, R.~H.}} \yr{1967}  \at{Intermittency in the very small
  scales of turbulence}.  \jt{Phys. Fluids}  \bvol{10},  \pg{2080--2082}.

\bibitem[Krommes(2008)]{krommes_08}
{\sc \au{Krommes, J.~A.}} \yr{2008}  \at{The remarkable similarity between the
  scaling of kurtosis with squared skewness for {TORPEX} density fluctuations
  and sea-surface temperature fluctuations}.  \jt{Phys. Plasmas}  \bvol{15},
  \pg{030703}.

\bibitem[Labit {\em et~al.\/}(2007)Labit, Furno, Fasoli, Diallo, M\"uller,
  Plyushchev, Podest\`a \& Poli]{labit_07}
{\sc \au{Labit, B.}, \au{Furno, I.}, \au{Fasoli, A.}, \au{Diallo, A.},
  \au{M\"uller, S.}, \au{Plyushchev, G.}, \au{Podest\`a, M.} \& \au{Poli, F.}}
  \yr{2007}  \at{Universal statistical properties of drift-interchange
  turbulence in {TORPEX} plasmas}.  \jt{Phys. Rev. Lett.}  \bvol{98},
  \pg{255002}.

\bibitem[Lenschow {\em et~al.\/}(1994)Lenschow, Mann \&
  Kristensen]{lenschow_94}
{\sc \au{Lenschow, D.}, \au{Mann, J.} \& \au{Kristensen, L.}} \yr{1994}
  \at{How long is long enough when measuring fluxes and other turbulence
  statistics?}  \jt{J. Atm. Oc. Tech.}  \bvol{11},  \pg{661--673}.

\bibitem[Losada {\em et~al.\/}(2023)Losada, Theodorsen \& Garcia]{losada_23}
{\sc \au{Losada, J.~M.}, \au{Theodorsen, A.} \& \au{Garcia, O.~E.}} \yr{2023}
  \at{Stochastic modeling of blob-like plasma filaments in the scrape-off
  layer: Theoretical foundation}.  \jt{Phys. Plasmas}  \bvol{30},  \pg{042518}.

\bibitem[Lovejoy {\em et~al.\/}(2001)Lovejoy, Schertzer \& Stanway]{lovejoy_01}
{\sc \au{Lovejoy, S.}, \au{Schertzer, D.} \& \au{Stanway, J.~D.}} \yr{2001}
  \at{Direct evidence of multifractal atmospheric cascades from planetary
  scales down to 1 km}.  \jt{Phys. Rev. Lett.}  \bvol{86},  \pg{5200--5203}.

\bibitem[Mallet {\em et~al.\/}(2017)Mallet, Schekochihin \&
  Chandran]{mallet_17j}
{\sc \au{Mallet, A.}, \au{Schekochihin, A.~A.} \& \au{Chandran, B.~D.}}
  \yr{2017}  \at{Disruption of alfv\'enic turbulence by magnetic reconnection
  in a collisionless plasma}.  \jt{JPP}  \bvol{466},  \pg{1--22}.

\bibitem[Marino {\em et~al.\/}(2022)Marino, Feraco, Primavera, Pumir, Pouquet
  \& Rosenberg]{marino_22}
{\sc \au{Marino, R.}, \au{Feraco, F.}, \au{Primavera, L.}, \au{Pumir, A.},
  \au{Pouquet, A.} \& \au{Rosenberg, D.}} \yr{2022}  \at{Turbulence generation
  by large-scale extreme drafts and the modulation of local energy dissipation
  in stably stratified geophysical flows}.  \jt{Phys. Rev. F}  \bvol{7},
  \pg{033801}.

\bibitem[Marino \& Sorriso-Valvo(2023)]{marino_23r}
{\sc \au{Marino, R.} \& \au{Sorriso-Valvo, L.}} \yr{2023}  \at{Scaling laws for
  the energy transfer in space plasma turbulence}.  \jt{Physics Reports}
  \bvol{1006},  \pg{1--144}.

\bibitem[Matthaeus {\em et~al.\/}(2015)Matthaeus, Wan, Servidio, Greco, Osman,
  Oughton \& Dmitruk]{matthaeus_15}
{\sc \au{Matthaeus, W.~H.}, \au{Wan, M.}, \au{Servidio, S.}, \au{Greco, A.},
  \au{Osman, K.~T.}, \au{Oughton, S.} \& \au{Dmitruk, P.}} \yr{2015}
  \at{Intermittency, nonlinear dynamics and dissipation in the solar wind and
  astrophysical plasmas}.  \jt{Phil. Trans. R. Soc. A}  \bvol{373},
  \pg{20140154}.

\bibitem[Meneguzzi {\em et~al.\/}(1996)Meneguzzi, Politano, Pouquet \&
  Zolver]{meneguzzi_96}
{\sc \au{Meneguzzi, M.}, \au{Politano, H.}, \au{Pouquet, A.} \& \au{Zolver,
  M.}} \yr{1996}  \at{A sparse-mode spectral method for the simulations of
  turbulent flows}.  \jt{J. Comp. Phys.}  \bvol{123},  \pg{32--44}.

\bibitem[Mezaoui {\em et~al.\/}(2014)Mezaoui, Hamza \&
  Jayachandran]{mezaoui_14}
{\sc \au{Mezaoui, H.}, \au{Hamza, A.} \& \au{Jayachandran, P.}} \yr{2014}
  \at{Investigating high-latitude ionospheric turbulence using global
  positioning system data}.  \jt{Geophys. Res. Lett.}  \bvol{41},
  \pg{6570--6576}.

\bibitem[Miranda {\em et~al.\/}(2018)Miranda, Schelin, Chian \&
  Ferreira]{miranda_18}
{\sc \au{Miranda, R.~A.}, \au{Schelin, A.~B.}, \au{Chian, A. C.-L.} \&
  \au{Ferreira, J.~L.}} \yr{2018}  \at{Non-{G}aussianity and cross-scale
  coupling in interplanetary magnetic field turbulence during a rope-rope
  magnetic reconnection event}.  \jt{Ann. Geophys.}  \bvol{36},  \pg{497--507}.

\bibitem[Nazarenko \& Schekochihin(2011)]{nazarenko_11}
{\sc \au{Nazarenko, S.} \& \au{Schekochihin, S.}} \yr{2011}  \at{Critical
  balance in magnetohydrodynamic, rotating and stratified turbulence: towards a
  universal scaling conjecture}.  \jt{J. Fluid Mech.}  \bvol{677},
  \pg{134--153}.

\bibitem[Petoukhov {\em et~al.\/}(2008)Petoukhov, Eliseev, Klein \&
  Oesterle]{petoukhov_08}
{\sc \au{Petoukhov, V.}, \au{Eliseev, A.}, \au{Klein, R.} \& \au{Oesterle, H.}}
  \yr{2008}  \at{On statistics of the free-troposphere synoptic component: an
  evaluation of skewnesses and mixed third-order moments contribution to the
  synoptic-scale dynamics and fluxes of heat and humidity}.  \jt{Tellus}
  \bvol{60A},  \pg{11--31}.

\bibitem[Politano \& Pouquet(1995)]{politano_95pr}
{\sc \au{Politano, H.} \& \au{Pouquet, A.}} \yr{1995}  \at{Model of
  intermittency in magnetohydrodynamic turbulence}.  \jt{Phys. Rev. E}
  \bvol{52},  \pg{636--641}.

\bibitem[Politano {\em et~al.\/}(1995)Politano, Pouquet \&
  Sulem]{politano_95pp}
{\sc \au{Politano, H.}, \au{Pouquet, A.} \& \au{Sulem, P.~L.}} \yr{1995}
  \at{Current and vorticity dynamics in three--dimensional turbulence}.
  \jt{Phys. Plasmas}  \bvol{2},  \pg{2931--2939}.

\bibitem[Ponty {\em et~al.\/}(2007)Ponty, Laval, Dubrulle, Daviaud \&
  Pinton]{ponty_07p}
{\sc \au{Ponty, Y.}, \au{Laval, J.}, \au{Dubrulle, B.}, \au{Daviaud, F.} \&
  \au{Pinton, J.-F.}} \yr{2007}  \at{Subcritical dynamo bifurcation in the
  {T}aylor-{G}reen flow}.  \jt{Phys. Rev. Lett.}  \bvol{99},  \pg{224501}.

\bibitem[Ponty {\em et~al.\/}(2005)Ponty, Mininni, Montgomery, Pinton, Politano
  \& Pouquet]{ponty_05}
{\sc \au{Ponty, Y.}, \au{Mininni, P.~D.}, \au{Montgomery, D.}, \au{Pinton,
  J.-F.}, \au{Politano, H.} \& \au{Pouquet, A.}} \yr{2005}  \at{Critical
  magnetic {R}eynolds number for dynamo action as a function of magnetic
  {Prandtl} number}.  \jt{Phys. Rev. Lett.}  \bvol{94},  \pg{164502}.

\bibitem[Pouquet {\em et~al.\/}(2025)Pouquet, Marino, Politano, Ponty \&
  Rosenberg]{pouquet_24}
{\sc \au{Pouquet, A.}, \au{Marino, R.}, \au{Politano, H.}, \au{Ponty, Y.} \&
  \au{Rosenberg, D.}} \yr{2025}  \at{Intermittency in fluid and {MHD}
  turbulence analyzed through the prism of moment scaling predictions of
  multifractal models}.  \jt{Nonlin. Proc. Geophys.}  \bvol{Submitted}.

\bibitem[Pouquet {\em et~al.\/}(2023)Pouquet, Rosenberg, Marino \&
  Mininni]{pouquet_23a}
{\sc \au{Pouquet, A.}, \au{Rosenberg, D.}, \au{Marino, R.} \& \au{Mininni, P.}}
  \yr{2023}  \at{Intermittency scaling for mixing and dissipation in rotating
  stratified turbulence at the edge of instability}.  \jt{Atmosphere}
  \bvol{14},  \pg{01375}.

\bibitem[Rafner {\em et~al.\/}(2021)Rafner, Grujic, Bach, B{\ae}rentzen,
  Gervang, Jia \& Misztal]{rafner_21}
{\sc \au{Rafner, J.}, \au{Grujic, Z.}, \au{Bach, C.}, \au{B{\ae}rentzen, J.},
  \au{Gervang, B.}, \au{Jia, R.and~Leinweber, S.} \& \au{Misztal,
  M.and~Sherson, J.}} \yr{2021}  \at{Geometry of turbulent dissipation and the
  navier-stokes regularity problem}.  \jt{Scientific Rep.}  \bvol{11},
  \pg{8824}.

\bibitem[Rincon(2019)]{rincon_dynamo_2019}
{\sc \au{Rincon, F.}} \yr{2019}  \at{Dynamo theories}.  \jt{J. Plasma Phys.}
  \bvol{85}~(4).

\bibitem[Sattin {\em et~al.\/}(2009)Sattin, Agostini, Scarin, Vianello,
  Cavazzana, Marrelli, Serianni, Zweben, Maqueda, Yagi, Sakakita, Koguchi,
  Kiyama, Hirano \& Terry]{sattin_09}
{\sc \au{Sattin, F.}, \au{Agostini, M.}, \au{Scarin, P.}, \au{Vianello, N.},
  \au{Cavazzana, R.}, \au{Marrelli, L.}, \au{Serianni, G.}, \au{Zweben, S.~J.},
  \au{Maqueda, R.}, \au{Yagi, Y.}, \au{Sakakita, H.}, \au{Koguchi, H.},
  \au{Kiyama, S.}, \au{Hirano, Y.} \& \au{Terry, J.}} \yr{2009}  \at{On the
  statistics of edge fluctuations: comparative study between various fusion
  devices}.  \jt{Plasma Phys. Control. Fusion}  \bvol{51},  \pg{055013}.

\bibitem[Schekochihin(2022)]{schekochihin_22}
{\sc \au{Schekochihin, A.~A.}} \yr{2022}  \at{{MHD} turbulence: A biased
  review}.  \jt{J. Plasma Phys.}  \bvol{88},  \pg{155880501}.

\bibitem[She \& L\'ev\^eque(1994)]{she_94}
{\sc \au{She, Z.} \& \au{L\'ev\^eque, E.}} \yr{1994}  \at{Universal scaling
  laws in fully developed turbulence}.  \jt{Phys. Rev. Lett.}  \bvol{72},
  \pg{336--339}.

\bibitem[Sladkomedova {\em et~al.\/}(2023)Sladkomedova, Cziegler, Field,
  Schekochihin \& Ivanov]{sladkomedova_23}
{\sc \au{Sladkomedova, A.}, \au{Cziegler, I.}, \au{Field, A.~R.},
  \au{Schekochihin, A.~A.} \& \au{Ivanov, P.~G.}} \yr{2023}  \at{Intermittency
  of density fluctuations and zonal-flow generation in {MAST} edge plasmas}.
  \jt{J. Plasma Phys.}  \bvol{89},  \pg{905890614 , 1--35}.

\bibitem[Sorriso-Valvo {\em et~al.\/}(2018)Sorriso-Valvo, Carbone, Perri,
  Greco, Marino \& Bruno]{sorriso_18}
{\sc \au{Sorriso-Valvo, L.}, \au{Carbone, F.}, \au{Perri, S.}, \au{Greco, A.},
  \au{Marino, R.} \& \au{Bruno, R.}} \yr{2018}  \at{On the statistical
  properties of turbulent energy transfer rate in the inner heliosphere}.
  \jt{Solar Phys.}  \bvol{293},  \pg{10}.

\bibitem[Sura \& Sardeshmukh(2008)]{sura_08}
{\sc \au{Sura, P.} \& \au{Sardeshmukh, P.~D.}} \yr{2008}  \at{A global view of
  non-gaussian {SST} variability}.  \jt{J. Phys. Oceano.}  \bvol{38},
  \pg{639--647}.

\bibitem[Sweet {\em et~al.\/}(2001)Sweet, Ott, Antonsen, Lathrop \&
  Finn]{sweet_01}
{\sc \au{Sweet, D.}, \au{Ott, E.}, \au{Antonsen, T.~M.}, \au{Lathrop, D.~P.} \&
  \au{Finn, J.~M.}} \yr{2001}  \at{Blowout bifurcations and the onset of
  magnetic dynamo action}.  \jt{Phys. Plasmas}  \bvol{8},  \pg{1944--1952}.

\bibitem[Theodorsen {\em et~al.\/}(2017)Theodorsen, Garcia \&
  Rypdal]{theodorsen_17}
{\sc \au{Theodorsen, A.}, \au{Garcia, O.} \& \au{Rypdal, M.}} \yr{2017}
  \at{Statistical properties of a filtered {P}oisson process with additive
  random noise: distributions, correlations and moment estimation}.
  \jt{Physica Scripta}  \bvol{92},  \pg{054002}.

\bibitem[Wan {\em et~al.\/}(2010)Wan, Oughton, Servidio \& Matthaeus]{wan_10}
{\sc \au{Wan, M.}, \au{Oughton, S.}, \au{Servidio, S.} \& \au{Matthaeus,
  W.~H.}} \yr{2010}  \at{On the accuracy of simulations of turbulence}.
  \jt{Phys. Plasmas}  \bvol{17},  \pg{082308}.

\bibitem[Wu {\em et~al.\/}(2023)Wu, Huang, Wang, Yuan, He \& Yang]{hu-hh_23}
{\sc \au{Wu, H.}, \au{Huang, S.}, \au{Wang, X.}, \au{Yuan, Z.}, \au{He, J.} \&
  \au{Yang, L.}} \yr{2023}  \at{Intermittency of magnetic discontinuities in
  the near-sun solar wind turbulence}.  \jt{Astrophys. J. Lett.}  \bvol{947},
  \pg{L22}.

\bibitem[Yeung {\em et~al.\/}(2015)Yeung, Zhai \& Sreenivasan]{yeung_15}
{\sc \au{Yeung, P.~K.}, \au{Zhai, X.~M.} \& \au{Sreenivasan, K.~R.}} \yr{2015}
  \at{Extreme events in computational turbulence}.  \jt{Proc. Nat. Acad. Sci.}
  \bvol{112},  \pg{12633--12638}.

\bibitem[Zhdankin {\em et~al.\/}(2017)Zhdankin, Boldyrev \& Mason]{zhdankin_17}
{\sc \au{Zhdankin, V.}, \au{Boldyrev, S.} \& \au{Mason, J.}} \yr{2017}
  \at{Influence of a large-scale field on energy dissipation in
  magnetohydrodynamic turbulence}.  \jt{MNRAS}  \bvol{468},  \pg{4025--4029}.

\bibitem[Zienicke {\em et~al.\/}(1998)Zienicke, Politano \&
  Pouquet]{zienicke_98}
{\sc \au{Zienicke, E.}, \au{Politano, H.} \& \au{Pouquet, A.}} \yr{1998}
  \at{Variable intensity of {L}agrangian chaos in the nonlinear dynamo
  problem}.  \jt{Phys. Rev. Lett.}  \bvol{81},  \pg{4640--4643}.

\end{thebibliography}
 
\end{document}